# Programming the Interactions of Collective Adaptive Systems by Relying on Attribute-based Communication

YEHIA ABD ALRAHMAN, IMT School for Advanced Studies, Lucca, Italy
ROCCO DE NICOLA, IMT School for Advanced Studies, Lucca, Italy
MICHELE LORETI, Università di Camerino, Camerino, Italy

Collective adaptive systems are new emerging computational systems consisting of a large number of interacting components and featuring complex behaviour. These systems are usually distributed, heterogeneous, decentralised and interdependent, and are operating in dynamic and possibly unpredictable environments. Finding ways to understand and design these systems and, most of all, to model the interactions of their components, is a difficult but important endeavour. In this article we propose a language-based approach for programming the interactions of collective-adaptive systems by relying on attribute-based communication; a paradigm that permits a group of partners to communicate by considering their run-time properties and capabilities. We introduce *AbC*, a foundational calculus for attribute-based communication and show how its linguistic primitives can be used to program a complex and sophisticated variant of the well-known problem of Stable Allocation in Content Delivery Networks. Also other interesting case studies, from the realm of collective-adaptive systems, are considered. We also illustrate the expressive power of attribute-based communication by showing the natural encoding of other existing communication paradigms into *AbC*.

## 1 INTRODUCTION

The ever increasing complexity of modern software systems has changed the perspective of software designers who now have to consider a broad range of new classes of systems, consisting of a large number of interacting components and featuring complex interaction mechanisms, e.g. *Software-Intensive Systems* [Broy 2006], *IoT Systems* [Kopetz 2011], and *Collective Adaptive Systems* (CAS) [Anderson et al. 2013; Ferscha 2015]. These systems are usually distributed, heterogeneous, decentralised and interdependent, and are operating more and more in dynamic and often unpredictable

This research has been partially supported by the European projects IP 257414 ASCENS and STReP 600708 QUANTICOL, and by the Italian project PRIN 2010LHT4KM CINA.

Author's addresses: Y.A. Alrahman and R. De Nicola, IMT School for Advanced Studies, Lucca, Italy; M. Loreti, Università di Camerino, Camerino, Italy;

environments. There exist different kinds of complexity in the development of software systems. Historically, as software systems grew larger, the focus shifted from the complexity of developing algorithms to the complexity of structuring large systems, and to the inevitable complexities of building distributed and concurrent systems [Wirsing and Holzl 2006]. We are now facing another level of complexity arising from the fact that systems have to operate in large, open and non-deterministic environments where the complexity of interaction is increased.

Most of the current communication models and programming frameworks still handle the interaction between distributed components by relying on their identities, like in the Actor model [Agha 1986], or by relying on the communication channels on which they are listening, like in point-to-point channel-based communication [Sangiorgi and Walker 2003], multicast with explicit addressing [Holbrook and Cheriton 1999], and broadcast communication [Prasad 1991]. These models and frameworks use names or addresses, that are totally independent from the run-time properties and capabilities of the interacting components, as first-class citizens to model the interaction. This makes it hard to program complex behaviours and coordination actions (reconfiguration, adaptation, opportunistic interactions, ...) that depend on the components' status rather than on their identities or addresses. The complexity of these models arises because they are *communication centric*; the behaviour mainly consists of a process implementing a specific communication protocol and the programmer has to manage the *channels* or the *addresses* on which components interact.

In our view, to deal with CAS, it is important to consider more *data centric* interaction primitives that abstract from the underlying communication infrastructure (i.e., infrastructure-agnostic) and relies on *anonymous one-to-many* interactions to send messages to potential receivers. In this article we study the impact of a new paradigm that permits a group of partners to interact by considering their properties and capabilities at run-time. The findings we report in this article have been triggered by our interest in CAS, and by the recent attempts to define appropriate linguistic primitives to deal with such systems, see e.g. Intelligent Ensembles [Krijt et al. 2017], FORMS [Weyns et al. 2012], Rainbow [Garlan et al. 2004], SOA-based [Calinescu 2008], Swarm of Cognitive Agents [Park and Tran 2012], UNDERSEA [Gerasimou et al. 2017], TOTA [Mamei and Zambonelli 2004], SCEL [De Nicola et al. 2014] and the calculi presented in [Alrahman et al. 2015] and [Viroli et al. 2013].

**Collective-Adaptive Systems (CAS)**

CAS [Anderson et al. 2013] consist of a large number of interacting components which combine their behaviours, by forming collectives, to achieve specific goals depending on their attributes, objectives, and functionalities. Decision-making in these systems is complex and components interaction may lead to unexpected behaviours. CAS are inherently scalable [Sanderson et al. 2015] and the boundaries between different CAS are fluid in the sense that components may enter or leave the collective at any time and may have different (potentially conflicting) objectives; so they need to dynamically adapt to their environmental conditions and contextual data. The need of engineering techniques to address the challenges of developing, integrating, and deploying CAS is advocated in [Sommerville et al. 2012]. Also the development of theoretical foundation for CAS is very important to understand their distinctive features [Cheng et al. 2009].

Clearly, the existing communication models do not scale with the high level of dynamicity of CAS, and, in our opinion, a change of perspective, that takes into account run-time properties, status, and capabilities of communicating systems, is on demand. The key concepts of CAS should be the basis for guiding the development of the new communication modalities.

In the following, we summarise the main concepts of CAS as reported in [Anderson et al. 2013; Cheng et al. 2009; Ferscha 2015; Sanderson et al. 2015]:

- *Distribution*: components are distributed on a shared environment and evolve independently, without any centralised control.
- *Awareness*: components are aware of their run-time status, characteristics, capabilities, and have some (partial) view of their surroundings.
- *Adaptivity*: components adapt their behaviours and their interaction policies in response to the changes in their contextual conditions and to the collected awareness data.
- *Interdependence*: any change in the shared environment might influence the behaviour of components.
- *Collaboration*: components collaborate and combine their behaviours to achieve system-level goals in response to changes in the environment or because of their predefined roles.
- *Anonymity*: components are able to communicate and exchange data without previously knowing the existence of each other; for example, the identity of a service provider is not relevant, only its features and its ability to provide the service are important.
- *Scalability and Open-endedness*: components may join or leave a system without disturbing the overall system behaviour; senders emit messages without being aware of the presence of receivers and receivers do not rely on specific senders.

**Attribute-based Communication**

In order to capture the above mentioned concepts, we have developed *attribute-based communication*, a communication paradigm that permits a group of partners to communicate by considering the predicates over the attributes they expose. Communication takes place anonymously in an implicit multicast fashion without a prior agreement between the communicating partners. Anonymity of interaction allows programmers to secure scalability, dynamicity, and open-endedness more easily. Sending operations are non-blocking while receiving operations are blocking. This breaks synchronisation dependencies between interacting partners, and permits modelling systems where communicating partners can enter or leave a group at any time without endangering the overall behaviour. Moreover, attributes make it easy to model awareness by locally reading the values of the attributes that may represent either the component status (e.g., the battery level of a robot) or the external environment (e.g., the external humidity). Groups are dynamically formed at the time of interaction by considering the interested receiving components that satisfy sender's predicates, and any run-time changes of attribute values allow opportunistic interactions between components. By parameterising the interaction predicates with local attributes, groups can be implicitly changed and adaptation is naturally captured. Security mechanisms can be placed on top of the attribute-based framework following standard approaches.

Modeling opportunistic interaction in classical communication paradigms like channel-based communication, e.g., $\pi$-calculus [Milner et al. 1992], is definitely more challenging because components have to agree on specific names or channels to interact. Channels have no connection with the component attributes or characteristics; they are specified as addresses where the exchange should happen. Names and channels are static and changing them locally at run-time requires explicit communication and intensive use of scoping mechanisms which do affect readability and compositionality of programs.

It is also worth mentioning that Multi-Agent System (MAS) [Wooldridge 2000] is usually used for modeling distributed and adaptive systems, e.g., [Leitão et al. 2012; Lepuschitz et al. 2013]. Since MAS originates as a sub-field of distributed artificial intelligence where logic is the reference formal framework, MAS models provide little support to formally validate an implementation against its specifications [Viroli and Omicini 2005]. Logics, however, provide powerful tools to declaratively model the features of complex systems as MAS [Wooldridge 2000]. In our approach we provide an operational model that precisely describes the step-by-step admissible evolutions of the system. Operational models ensure more straightforwardly that implementations are correct with respect to their specifications and based on the evolution rules, system properties can be verified [Viroli and Omicini 2005]. Different attempts have been carried out to bridge the gap between MAS and process algebra approaches, see [Bergenti et al. 2003; de Boer et al. 2005; Kinny 2005], however this is out the scope of this article.

We would also like to stress that an attribute-based system is more than just the result of the parallel composition of its interacting components; it also takes into account the environment where components are executed. Indeed, the environment has a great impact on components behaviours and allows modelling interdependence, i.e., a situation where components influence each other unintentionally. For example, in an ant foraging system [Jackson and Ratnieks 2006], when an ant disposes pheromone in the shared space to keep track of its way back home, it influences other ants behaviour as they are programmed to follow traces of pheromone with higher concentration. In this way, the ant unintentionally influences the behaviour of the other ants by only modifying the environment.

In this article we introduce *AbC*, a calculus comprising a minimal set of primitives that permits attribute-based communication and is the outcome of different attempts towards modeling the interaction of CAS scenarios. From our experience, we learnt that any attribute-based paradigm, tailored for modeling the interaction of CAS systems, should at least provide support for the following notions, that are the key ingredients of *AbC*:

- *Attribute Environment*: to provide a collection of attributes whose values represent the status of a component. These values can be used to influence the behaviour of a component at run-time.
- *Attribute-based send and receive operations*: to establish the communication links between different components based on the satisfaction of predicates over components attributes.
- *Attribute update operation*: to change attribute values based on contextual conditions and adapt the behaviour of a component accordingly.
- *Awareness construct*: to collect awareness data and take decisions based on the changes in the attribute environment.

In this article, we focus on a language-based approach to program the interactions of CAS and concentrate on the linguistic constructs of the *AbC* calculus and on their use. Other more theoretical aspects, concerned with modeling, with primitives for abstraction, with behavioural theories, and with equational laws, can be found in [Alrahman et al. 2016a, 2017a].

An *AbC* system is rendered as a set of parallel components, each equipped with a set of attributes whose values can be modified by internal actions. Communication actions (both send and receive) are decorated with predicates over attributes that partners have to satisfy to make the interaction possible. Thus, communication takes place in an implicit multicast fashion, and communication partners are selected by relying on predicates over the attributes in their interfaces. Unlike IP multicast [Holbrook and Cheriton 1999] where the reference address of the group is explicitly included in the message, *AbC* components are unaware of the existence of each other and they receive messages only if they mutually satisfy the requirements of each other, otherwise messages are discarded.

Our main goal is to permit the construction of high-level and formally verifiable communication APIs for CAS based on the *AbC* calculus. These APIs can then be used to reason and verify properties about the interactions of the actual deployed systems by relying on the operational semantics of *AbC*. For example, we developed APIs for Java [Alrahman et al. 2016b], for Go [Alrahman et al. 2017b], and for Erlang [De Nicola et al. 2017]. The actual implementations of these APIs fully rely on the formal semantics of *AbC*. We also provided a one-to-one correspondence between the *AbC* primitives and the programming constructs of the above mentioned APIs. The direct correspondence permits exploiting the *AbC* linguistic primitives to program the interaction of CAS applications in different host languages as required by the application domain. At the same time, it enhances the confidence on the behaviour of programs, written by using these APIs, after they have been analysed via formal methods, which is made possible by relying on the operational semantics of the *AbC* calculus.

*Contributions.* The main contribution of this article is the introduction of a language-based approach for programming the interactions of collective-adaptive systems. In Section 2, we introduce the *AbC* calculus. To capture the reader's intuition, we present the main features of *AbC* in a step-by-step fashion by using a distributed variant of the *Graph Colouring Problem* [Jensen and Toft 1995] as a running example. In Section 2.3, we illustrate the expressive power of attribute-based communication by showing the natural encoding of other communication paradigms. In the rest of the article, we exploit the interaction primitives of the *AbC* calculus to show how complex and interesting case studies, from the realm of collective-adaptive systems, can be programmed in an intuitive way. More specifically, in Section 3, we show how to program a more complex and sophisticated variant of the well-known problem of Stable Allocation in Content Delivery Networks [Maggs and Sitaraman 2015]. This variant is designed to cover most of CAS features in a single case study with a well-understood system-level goal. Additional case studies are presented in the appendices A.1 and A.2. In A.1, we show how to program a crowd steering scenario while in Section A.2, we show how to program swarm of robotics, performing a rescuing mission in a disaster arena.

## 2 THE ABC CALCULUS

In this section we first present the syntax of *AbC*, we discuss its formal operational semantics, and then we provide an evidence of its expressive power.

### 2.1 Syntax of the *AbC* Calculus

The syntax of *AbC* is reported in Table 1. The top-level entities of the calculus are *components* ($C$). A component, $\Gamma:_I P$, is a process $P$ associated with an *attribute environment* $\Gamma$, and an *interface* $I$. An *attribute environment* $\Gamma: \mathcal{A} \rightharpoonup \mathcal{V}$ is a partial map from attribute identifiers[1] $a \in \mathcal{A}$ to values $v \in \mathcal{V}$ where $\mathcal{A} \cap \mathcal{V} = \emptyset$. A value could be a number, a name (string), a tuple, etc. An *interface* $I \subseteq \mathcal{A}$ consists of a set of *attributes* that are exposed by a component to control the interactions with other components. We will refer to the attributes in $I$ as *public attributes*, and to those in $dom(\Gamma) - I$ as *private attributes*. Complex components are built by using parallel operator $C_1 \| C_2$.

*Example 2.1 (Distributed Graph Colouring Scenario).* We model a distributed variant of the well known *Graph Colouring Problem* [Jensen and Toft 1995] using *AbC* constructs. We render the problem as a typical CAS scenario where a collective of agents, executing the same code, collaborate to achieve a system-level goal without any centralised control.

The problem consists of colouring the vertices of a graph in such a way that no two vertices sharing an edge have the same colour. Formally, we have a set of *n* vertices, each of which is identified by an *id* (an integer in our model). A set of *neighbours* $N_i$ is also associated with each vertex $i$, where $j \in N_i$ if and only if $i \in N_j$. Our goal is to assign to each vertex $i$ a colour $c_i$ such that for each $j \in N_i$, $c_j \neq c_i$.

---
[1]In the rest of this article, we shall occasionally use the term "attribute" instead of "attribute identifier".

| | |
|---|---|
| (Components) | $C ::= \Gamma :_I P \mid C_1 \| C_2$ |
| (Processes) | $P ::= 0 \mid \Pi(\tilde{x}).U \mid (\tilde{E})@\Pi.U \mid \langle\Pi\rangle P \mid P_1 + P_2 \mid P_1 \mid P_2 \mid K$ |
| (Updates) | $U ::= [a := E]U \mid P$ |
| (Predicates) | $\Pi ::= \text{tt} \mid \text{ff} \mid p(\tilde{E}) \mid \Pi_1 \wedge \Pi_2 \mid \Pi_1 \vee \Pi_2 \mid \neg\Pi$ |
| (Expressions) | $E ::= v \mid x \mid a \mid \text{this}.a \mid op(\tilde{E})$ |

Table 1. The syntax of the *AbC* calculus

Each vertex is modelled in *AbC* as a component of the form $C_i = \Gamma_i :_{\{\text{id},\text{N}\}} P_C$. Public attributes id and N are used to represent the vertex *id* and the set of *neighbours* N, respectively. The overall system is defined as the parallel composition of existing components (i.e., $C_1 \| C_2 \|, \ldots, \| C_n$).  □

A *process* can be the *inactive* process 0, an *action-prefixed* process, $act.U$, where $act$ is a communication action and $U$ is a process possibly preceded by an *attribute update*, a *context aware* $\langle\Pi\rangle P$ process, a *nodeterministic choice* between two processes $P_1 + P_2$, a *parallel composition* of two processes $P_1 | P_2$, or a process call with a unique identifier $K$ used in process definition $K \triangleq P$. All of these operators will now be described below. We start by explaining what we mean by expressions and predicates, then we continue by describing the actual operations on processes.

An *expression* $E$ is built from constant values $v \in \mathcal{V}$, variables $x$, attribute identifiers $a$, a reference to the value of $a$ (this.$a$) in the component that is executing the code, or through a standard operators $op(\tilde{E})$[2]. The evaluation of expression $E$ under $\Gamma$ is denoted by $[\![E]\!]_\Gamma$. The definition of $[\![\cdot]\!]_\Gamma$ is standard, the only interesting cases are $[\![a]\!]_\Gamma = [\![\text{this}.a]\!]_\Gamma = \Gamma(a)$.

A *predicate* $\Pi$ is built from boolean constants, tt and ff, and from an *atomic predicate* $p(\tilde{E})$ by using standard boolean operators ($\neg$, $\wedge$ and $\vee$). The precise set of atomic predicates is not detailed here; we only assume that it contains basic binary relations like $>$, $<$, $\leq$, $\geq$, $=$, and the predicates $\in$ and $\notin$. In what follows, we shall use the notation $\{\Pi\}_\Gamma$ to indicate the *closure* of a predicate $\Pi$ under the attribute environment $\Gamma$. The closure is also a predicate $\Pi'$ obtained from $\Pi$ by replacing the occurrences of the expression this.$a$ with its value $\Gamma(a)$.

The *attribute-based output* $(\tilde{E})@\Pi$ is used to send the evaluation of the sequence of expressions $\tilde{E}$ to the components whose attributes satisfy the predicate $\Pi$.

The *attribute-based input* $\Pi(\tilde{x})$ is used to receive messages from a component satisfying predicate $\Pi$; the sequence $\tilde{x}$ acts as a placeholder for received values. The action $\Pi(\tilde{x})$ acts as a binder for names, e.g., $\tilde{x}$ in $\Pi(\tilde{x}).U$. We will say that a name $x$ is *bound* when it occurs under the scope of an input action while it is *free* when it is not bound. We use $bn(P)$ and $fn(P)$ to denote the set of bound and free names of $P$, respectively. We use $fv(P)$ to denote the set of free process variables of $P$. We use $x, y, \ldots$ to range over names while $X, Y, \ldots$ to range over process variables. Notice that names are used as placeholders for values while process variables are used as placeholders for processes. Our processes are *closed*, i.e. without free process variables ($fv(P) = \emptyset$) because *AbC* components can only exchange values, but not code.

---

[2]For the sake of simplicity, we omit the specific syntax of operators used to build expressions and use $\tilde{E}$ to denote sequences of expressions.

Predicates, used in communication actions, can also refer to variable names in $\tilde{x}$ and the received values can be used to check whether specific conditions are satisfied. For instance, the action

$$((x = \text{``}try\text{''}) \land (\text{this.id} > \text{id}) \land (\text{this.round} = z))(x, y, z)$$

can be used to receive a message of the form $(\text{``}try\text{''}, c, r)$ where the value received on $z$ is equal to this.round and the value of the interface attribute id of the sending component is less than this.id. Thus, the predicate can be used to check either the received values or the values of the sending component interface. A predicate can also refer to *local* attributes of components. Thus, an action like

$$(\text{``}try\text{''}, c, r)@(\text{this.id} \in \text{N})$$

can be used to send the message $(\text{``}try\text{''}, c, r)$ to all components whose attribute N contains this.id.

An *attribute update*, $[a := E]$, is used to assign the result of the evaluation of $E$ to the attribute identifier $a$. The syntax is devised in such a way that sequences of updates are only possible after communication actions. Actually, updates can be viewed as side effects of interactions. It should be noted that the execution of a communication action and the following update(s) is atomic. This possibility allows components to modify their attribute values and thus triggering new behaviours in response to collected contextual data.

The *awareness construct*, $\langle\Pi\rangle P$, is used to trigger new behaviours (i.e., $P$) when the status of a component is changed (i.e., $\Pi \models \Gamma$). It blocks the execution of $P$ until predicate $\Pi$ satisfies the attribute environment.

The *parallel operator*, $P|Q$, models the interleaving between co-located processes, i.e., processes residing within the same component.

The *choice operator*, $P + Q$, indicates a nondeterministic choice among $P$ and $Q$.

Other process operators can be defined as *macros* in *AbC*. Indeed, we will use the following derived operators:

$$\textbf{if } \Pi \textbf{ then } P_1 \textbf{ else } P_2 \triangleq \langle\Pi\rangle P_1 + \langle\neg\Pi\rangle P_2 \tag{1}$$

$$\textbf{let } x = E \textbf{ in } P \triangleq P[E/x] \tag{2}$$

$$\textbf{set}(a, E)P \triangleq ()@\text{ff}.[a := E]P \tag{3}$$

$$[a_1 := E_1,\ a_2 := E_2,\ \ldots,\ a_n := E_n]P \triangleq [a_1 := E_1][a_2 := E_2]\ldots[a_n := E_n]P \tag{4}$$

*Example 2.2.* In Example 2.1 we introduced the structure of the components modelling the vertices in the considered scenario. Here, we show how the behaviour of our components can be programmed to assign a *colour* (an integer) to each of them while avoiding that two neighbours get the same colour.

The proposed algorithm consists of a sequence of rounds for colour selection that goes on until the specified goal is reached. At the end of each round at least one component is assigned a colour.

Components use messages of the form $(\text{``}try\text{''}, c, r)$ to inform their neighbours that at round $r$ they want to select colour $c$ and messages of the form $(\text{``}done\text{''}, c, r)$ to communicate that colour $c$ has been definitely chosen at the end of round $r$. At the beginning of a round, each component $i$ selects a colour and sends a *try*-message to all components in $N_i$. Component $i$ also collects *try*-messages from its neighbours. The selected colour is assigned to a component only if it has the greatest id among those that have selected the same colour in that round. After the assignment, a *done*-message (associated with the current round) is sent to neighbours. A new round starts when a message, associated with a round $r$ such that this.round $< r$, is received.

This algorithm can be implemented in *AbC* by using four processes, *F* for forwarding *try*-messages to neighbours, *T* for handling *try*-messages, *D* for handling *done*-messages, and *A* for assigning a final colour. Process $P_C$ of Example 2.1 is now defined as the parallel composition of these four processes: $P_C = F \mid T \mid D \mid A$.

The following *private attributes*, local to each component, are used to control the progress of our algorithm: round, done, assigned, used, counter, send and constraints. The attribute "round" stores the current round while "constraints" and "used" are sets, registering the colours used by neighbours. The attribute "counter" counts the number of messages collected by a component while "send" is used to enable/disable forwarding of messages to neighbours. The private attributes of each component are all initialised with the following values: round = done = 0, constraints = used = $\emptyset$, send = tt, and assigned = ff.

In process *F* reported below, when the value of attribute send becomes tt, a new colour is selected, and a message containing this colour and the current round is sent to all the components having this.id as neighbour. The new colour is the smallest colour that has not yet been selected by neighbours, that is $\min\{i \notin \text{this}.used\}$. The guard ¬assigned is used to make sure that components with assigned colours do not take part in the selection anymore.

$$F \triangleq \langle \text{send} \wedge \neg\text{assigned} \rangle$$
$$\textbf{set}(\text{colour}, \min\{i \notin \text{this}.used\})$$
$$(\text{``}try\text{''}, \text{this.colour}, \text{this.round})@(\text{this}.id \in \mathsf{N}).[\text{send} := \text{ff}]F$$

Process *T*, reported below, receives messages of the form ("*try*", *c*, *r*). If $r = \text{this}.round$, as in the first two branches, then the received message has been originated by another component performing the same round of the algorithm. The first branch is executed when this.id > id, i.e., the sender has an *id* smaller than the *id* of the receiver. In this case, the message is ignored (there is no conflict), simply the counter of received messages (this.counter) is incremented. In the second branch, this.id < id, the received colour is recorded to check the presence of conflicts. The value of *y* is added to this.constraints and this.counter is incremented by 1.

If $r > \text{this}.round$, as in the last two branches, then the received message has been originated by a component executing a successive round and two possible alternatives are considered, this.id < id or this.id > id. In both cases, round is set to *r*, *send* and *counter* are updated accordingly, and this.constraints is set to the value of *y* if this.id < id.

$$T \triangleq \quad ((x = \text{``}try\text{''}) \wedge (\text{this}.id > id) \wedge (\text{this.round} = z))(x, y, z).$$
$$[\text{counter} := \text{counter} + 1]T$$

$$+ ((x = \text{``}try\text{''}) \wedge (\text{this}.id < id) \wedge (\text{this.round} = z))(x, y, z).$$
$$[\text{counter} := \text{counter} + 1, \text{constraints} := \text{constraints} \cup \{y\}]T$$

$$+ ((x = \text{``}try\text{''}) \wedge (\text{this}.id > id) \wedge (\text{this.round} < z))(x, y, z).$$
$$[\text{round} := z, \text{send} := \text{tt}, \text{counter} := 1, \text{constraints} := \emptyset]T$$

$$+ ((x = \text{``}try\text{''}) \wedge (\text{this}.id < id) \wedge (\text{this.round} < z))(x, y, z).$$
$$[\text{round} := z, \text{send} := \text{tt}, \text{counter} := 1, \text{constraints} := \{y\}]T$$

Process *D*, reported below, is used to receive messages of the form ("*done*", *c*, *r*). These are sent by components that have reached a final decision about their colour. When ("*done*", *c*, *r*) is received, we have that either $\text{this.round} \geq r$ or $\text{this.round} < r$. In the first case, the used colour is registered and the counter this.done is incremented. In the second case, private attributes are updated to indicate the startup of a new round.

$$D \triangleq \quad ((x = \text{``}done\text{''}) \wedge (\text{this.round} \geq z))(x, y, z).$$
$$[\text{done} := \text{done} + 1, \text{used} := \text{used} \cup \{y\}]D$$
$$+ ((x = \text{``}done\text{''}) \wedge (\text{this.round} < z))(x, y, z).$$
$$[\text{round} := z, \text{done} := \text{done} + 1, \text{constraints} := \emptyset,$$
$$\text{send} := \text{tt}, \text{counter} := 0, \text{used} := \text{used} \cup \{y\}]D$$

Process $A$, reported below, is used to manage the definitive selection of a colour and can only be executed when messages from all "pending" neighbours have been received ($\text{this.counter} = |\text{this.N}| - \text{this.done}$) and no conflict has been found ($\text{this.colour} \notin \text{this.used} \cup \text{this.constraints}$). When the above conditions are satisfied, message ("$done$", $\text{this.colour}, \text{this.round} + 1$) is sent to neighbours, the assigned attribute is set to true, and the process terminates.

$$A \triangleq \langle (\text{this.}counter = |\text{this.N}| - \text{this.done}) \wedge (\text{this.colour} \notin \text{this.constraints} \cup \text{this.used}) \rangle$$
$$(\text{``}done\text{''}, \text{this.colour}, \text{this.round} + 1)@(\text{this.id} \in \text{N}).[\text{assigned} := \text{tt}]0$$

□

*Remark.* Example 2.2 shows the main advantages of the attribute-based interaction. In essence, components are *infrastructure-agnostic*, i.e., they abstract from the underlying communication infrastructure and rely on *anonymous one-to-many* interaction pattern to communicate; this simplifies the design of component behaviours, because there is no need to manage the communication channels or the addresses on which components interact. The use of communication predicates, to derive the interaction between different components, permits programming *data centric* applications by taking into account the run-time features of the interacting components.

## 2.2 AbC Operational Semantics

In this section, we provide an overview of the operational semantics of *AbC* and use fragments of the Distributed Graph Colouring example to show how the semantics rules work. The operational semantics of *AbC* is based on two relations. The transition relation $\mapsto$ that describes the behavior of single components and the transition relation $\rightarrow$ that relies on $\mapsto$ and describes system behaviours.

*2.2.1 Operational semantics of components.* We use the transition relation $\mapsto \subseteq \text{Comp} \times \text{CLAB} \times \text{Comp}$ to define the local behavior of a component where Comp denotes the set of components and CLAB is the set of transition labels $\alpha$ generated by the following grammar:

$$\alpha ::= \lambda \quad | \quad \widetilde{\Gamma \triangleright \Pi(\tilde{v})} \qquad\qquad \lambda ::= \Gamma \triangleright \overline{\Pi}(\tilde{v}) \quad | \quad \Gamma \triangleright \Pi(\tilde{v})$$

The $\lambda$-labels are used to denote *AbC* output ($\Gamma \triangleright \overline{\Pi}(\tilde{v})$) and input ($\Gamma \triangleright \Pi(\tilde{v})$) actions. The former contains the sender's predicate $\Pi$, that specifies the expected communication partners, the transmitted values $\tilde{v}$ and the portion of the sender *attribute environment* $\Gamma$ that can be perceived by receivers. The latter is just the complementary label selected among all the possible ones that the receiver may accept. The $\alpha$-labels include an additional label $\widetilde{\Gamma \triangleright \Pi(\tilde{v})}$ to model the case where a component is not able to receive a message. As it will be seen later, this kind of *negative* labels is crucial to appropriately handle dynamic operators like choice and awareness.

$$\Gamma \models \text{tt for all } \Gamma$$
$$\Gamma \not\models \text{ff for all } \Gamma$$
$$\Gamma \models p(\tilde{E}) \text{ iff } [\![\tilde{E}]\!]_\Gamma \in [\![p]\!]_\Gamma$$
$$\Gamma \models \Pi_1 \wedge \Pi_2 \text{ iff } \Gamma \models \Pi_1 \text{ and } \Gamma \models \Pi_2$$
$$\Gamma \models \Pi_1 \vee \Pi_2 \text{ iff } \Gamma \models \Pi_1 \text{ or } \Gamma \models \Pi_2$$
$$\Gamma \models \neg\Pi \text{ iff not } \Gamma \models \Pi$$

Table 2. The predicate satisfaction

$$\frac{[\![\tilde{E}]\!]_\Gamma = \tilde{v} \quad \{\Pi_1\}_\Gamma = \Pi}{\Gamma:_I (\tilde{E})@\Pi_1.U \xmapsto{\Gamma\downarrow I \triangleright \overline{\Pi}(\tilde{v})} \{\!|\Gamma:_I U|\!\}} \text{ Brd} \qquad \frac{}{\Gamma:_I (\tilde{E})@\Pi.P \xmapsto{\widetilde{\Gamma'\triangleright\Pi'(\tilde{v})}} \Gamma:_I (\tilde{E})@\Pi.P} \text{ FBrd}$$

$$\frac{\Gamma' \models \{\Pi_1[\tilde{v}/\tilde{x}]\}_{\Gamma_1} \quad \Gamma_1 \downarrow I \models \Pi}{\Gamma_1:_I \Pi_1(\tilde{x}).U \xmapsto{\Gamma'\triangleright\Pi(\tilde{v})} \{\!|\Gamma_1:_I U[\tilde{v}/\tilde{x}]|\!\}} \text{ Rcv} \qquad \frac{\Gamma' \not\models \{\Pi[\tilde{v}/\tilde{x}]\}_\Gamma \vee \Gamma_1 \downarrow I \not\models \Pi'}{\Gamma_1:_I \Pi(\tilde{v}).U \xmapsto{\widetilde{\Gamma'\triangleright\Pi'(\tilde{v})}} \Gamma_1:_I \Pi(\tilde{v}).U} \text{ FRcv}$$

$$\frac{\Gamma \models \Pi \quad \Gamma:_I P \xmapsto{\lambda} \Gamma':_I P'}{\Gamma:_I \langle\Pi\rangle P \xmapsto{\lambda} \Gamma':_I P'} \text{ Aware} \qquad \frac{\Gamma \not\models \Pi}{\Gamma:_I \langle\Pi\rangle P \xmapsto{\widetilde{\Gamma'\triangleright\Pi'(\tilde{v})}} \Gamma:_I \langle\Pi\rangle P} \text{ FAware1}$$

$$\frac{\Gamma \models \Pi \quad \Gamma:_I P \xmapsto{\widetilde{\Gamma'\triangleright\Pi'(\tilde{v})}} \Gamma:_I P}{\Gamma:_I \langle\Pi\rangle P \xmapsto{\widetilde{\Gamma'\triangleright\Pi'(\tilde{v})}} \Gamma:_I \langle\Pi\rangle P} \text{ FAware2}$$

Table 3. Operational Semantics of Components (Part 1)

The transition relation $\mapsto$ is defined in Table 3 and Table 4 inductively on the syntax of Table 1. For each process operator we have two types of rules: one describing the actions a term can perform, the other showing how a component discards undesired input messages.

The behaviour of an *attribute-based output* is defined by rule BRD in Table 3. This rule states that when an output is executed, the sequence of expressions $\tilde{E}$ is evaluated, say to $\tilde{v}$, and the *closure* $\Pi$ of predicate $\Pi_1$ under $\Gamma$ is computed. Hence, these values are sent to other components together with $\Gamma \downarrow I$. This represents the portion of the *attribute environment* that can be perceived by the context and it is obtained from the local $\Gamma$ by limiting its domain to the attributes in the interface $I$:

$$(\Gamma \downarrow I)(a) = \begin{cases} \Gamma(a) & a \in I \\ \bot & \text{otherwise} \end{cases}$$

Afterwards, possible updates $U$, following the action, are applied. This is expressed in terms of a recursive function $\{\!|C|\!\}$ defined below:

$$\{\!|C|\!\} = \begin{cases} \{\!|\Gamma[a \mapsto [\![E]\!]_\Gamma]:_I U |\!\} & C \equiv \Gamma:_I [a := E]U \\ \Gamma:_I P & C \equiv \Gamma:_I P \end{cases}$$

where $\Gamma[a \mapsto v]$ denotes an attribute update such that $\Gamma[a \mapsto v](a') = \Gamma(a')$ if $a \neq a'$ and $v$ otherwise.

$$\frac{\Gamma:_I P_1 \stackrel{\lambda}{\mapsto} \Gamma':_I P_1'}{\Gamma:_I P_1 + P_2 \stackrel{\lambda}{\mapsto} \Gamma':_I P_1'} \text{SumL} \qquad \frac{\Gamma:_I P_2 \stackrel{\lambda}{\mapsto} \Gamma':_I P_2'}{\Gamma:_I P_1 + P_2 \stackrel{\lambda}{\mapsto} \Gamma':_I P_2'} \text{SumR}$$

$$\frac{\Gamma:_I P_1 \stackrel{\widetilde{\Gamma' \triangleright \Pi'(\tilde{v})}}{\longmapsto} \Gamma:_I P_1 \quad \Gamma:_I P_2 \stackrel{\widetilde{\Gamma' \triangleright \Pi'(\tilde{v})}}{\longmapsto} \Gamma:_I P_2}{\Gamma:_I P_1 + P_2 \stackrel{\widetilde{\Gamma' \triangleright \Pi'(\tilde{v})}}{\longmapsto} \Gamma:_I P_1 + P_2} \text{FSum}$$

$$\frac{\Gamma:_I P_1 \stackrel{\lambda}{\mapsto} \Gamma':_I P'}{\Gamma:_I P_1 \mid P_2 \stackrel{\lambda}{\mapsto} \Gamma':_I P' \mid P_2} \text{IntL} \qquad \frac{\Gamma:_I P_2 \stackrel{\lambda}{\mapsto} \Gamma':_I P'}{\Gamma:_I P_1 \mid P_2 \stackrel{\lambda}{\mapsto} \Gamma':_I P_1 \mid P'} \text{IntR}$$

$$\frac{\Gamma:_I P_1 \stackrel{\widetilde{\Gamma' \triangleright \Pi'(\tilde{v})}}{\longmapsto} \Gamma:_I P_1 \quad \Gamma:_I P_2 \stackrel{\widetilde{\Gamma' \triangleright \Pi'(\tilde{v})}}{\longmapsto} \Gamma:_I P_2}{\Gamma:_I P_1 \mid P_2 \stackrel{\widetilde{\Gamma' \triangleright \Pi'(\tilde{v})}}{\longmapsto} \Gamma:_I P_1 \mid P_2} \text{FInt}$$

$$\frac{\Gamma:_I P \stackrel{\lambda}{\mapsto} \Gamma':_I P' \quad K \triangleq P}{\Gamma:_I K \stackrel{\lambda}{\mapsto} \Gamma':_I P'} \text{Rec} \qquad \frac{\Gamma:_I P \stackrel{\widetilde{\Gamma' \triangleright \Pi'(\tilde{v})}}{\longmapsto} \Gamma:_I P \quad K \triangleq P}{\Gamma:_I K \stackrel{\widetilde{\Gamma' \triangleright \Pi'(\tilde{v})}}{\longmapsto} \Gamma:_I K} \text{FRec}$$

$$\frac{}{\Gamma:_I 0 \stackrel{\widetilde{\Gamma' \triangleright \Pi'(\tilde{v})}}{\longmapsto} \Gamma:_I 0} \text{FZero}$$

Table 4. Operational Semantics of Components (Part 2)

*Example 2.3.* In Example 2.2, an attribute-based output is used by components to communicate their colour. If we take component $C_1 = \Gamma_1 :_{\{id, N\}} F_1$, with $\Gamma_1 = \{id = 0, N = \{1, 2\}, \text{colour} = 0, \text{round} = 0, \ldots\}$ and $F_1 = (\text{"try"}, \text{this.colour}, \text{this.round})@(\text{this.id} \in N).[\text{this.send} := \mathit{false}]F$, it is easy to see that, by using rule BrD, the following transition can be derived:

$$C_1 \xrightarrow{\{id=0, N=\{1,2\}\} \triangleright \overline{0 \in N}(\text{"try"}, 0, 0)} \Gamma_1[\text{send} \mapsto \text{ff}] :_{\{id, N\}} F$$

Rule BrD is not sufficient to fully describe the behaviour of an output action; we need another rule (FBrD) to model the fact that all inputs are *discarded* in case only output actions are possible.

Rule Rcv governs the execution of input actions. It states that a message can be received when two *communication constraints* are satisfied: the local attribute environment restricted to interface $I$ ($\Gamma_1 \downarrow I$) satisfies $\Pi$, the predicate used by the sender to identify potential receivers; the sender environment $\Gamma'$ satisfies the receiving predicate $\{\Pi_1[\tilde{v}/\tilde{x}]\}_{\Gamma_1}$. When these two constraints are satisfied the input action is performed and the update $U$ is applied under the substitution $[\tilde{v}/\tilde{x}]$. The predicate satisfaction relation, $\models$, is reported in Table 2. Rule FRcv states that an input is *discarded* when the local attribute environment does not satisfy the *sender's predicate*, or the *receiving predicate* is not satisfied by the sender's environment.

*Example 2.4.* Let us consider the component $C_2 = \Gamma_1 :_{\{id,N\}} T'$ where $\Gamma_1(\text{id}) = 2$, $\Gamma_1(\text{round}) = 2$, $\Gamma_1(\text{counter}) = 0$ and

$$T' = ((x = \text{``}try\text{''}) \wedge (\text{this.id} > \text{id}) \wedge (\text{this.round} = z))(x,y,z).[\text{counter} := \text{counter} + 1]T$$

The following transition can be derived by using rules Rcv:

$$C_2 \xmapsto{\{id=1,\ldots\} \triangleright id \in \{0,2\}(\text{``}try\text{''},1,2)} \Gamma_1[\text{counter} \mapsto 1] :_{\{id,N\}} T$$

Instead, the following two negative transitions can be derived by using FRcv:

$$C_2 \xmapsto{\{id=3,\ldots\} \triangleright \widetilde{id \in \{0,2\}}(\text{``}try\text{''},1,2)} C_2$$

$$C_2 \xmapsto{\{id=1,\ldots\} \triangleright \widetilde{id \in \{0,3\}}(\text{``}try\text{''},1,3)} C_2$$

In the first transition the message is discarded because sender's id = 3 and predicate this.id > id is not satisfied. While in the second one message is discarded because *sender predicate* id $\in \{0,3\}$ is not satisfied by $\Gamma_1 \downarrow \{id, N\}$.

The behaviour of a component $\Gamma :_I \langle \Pi \rangle P$ is the same as of $\Gamma :_I P$ only when $\Gamma \models \Pi$, while the component is inactive when $\Gamma \not\models \Pi$. This is rendered by rules Aware, FAware1 and FAware2.

Rules SumL, SumR, and FSum describe behaviour of $\Gamma :_I P_1 + P_2$. Rules SumL and SumR are standard and just say that $\Gamma :_I P_1 + P_2$ behaves nondeterministically either like $\Gamma :_I P_1$ or like $\Gamma :_I P_2$. A message is *discarded* by $\Gamma :_I P_1 + P_2$ if and only if both $P_1$ and $P_2$ are not able to receive it. We can observe here that the presence of discarding rules is fundamental to prevent processes that cannot receive messages from evolving without performing actions. Thus *dynamic operators*, that are the ones *disappearing* after a transition like awareness and choice, persist after a message refusal.

The behaviour of the interleaving operator is described by rules IntL, IntR and FInt. The first two are standard process algebraic rules for parallel composition while the discarding rule FInt has a similar interpretation as of rule FSum: a message can be discarded only if both the parallel processes can discard it.

Finally, rules Rec, FRec and FZero are the standard rules for handling process definition and the inactive process. The latter states that process 0 always discards messages.

2.2.2 *Operational semantics of systems.* The behaviour of an *AbC* system is described by means of the transition relation $\rightarrow \subseteq \text{Comp} \times \text{SLAB} \times \text{Comp}$, where Comp denotes the set of components and SLAB is the set of transition labels $\lambda$ which are generated by the following grammar:

$$\lambda ::= \Gamma \triangleright \overline{\Pi}(\tilde{v}) \quad | \quad \Gamma \triangleright \Pi(\tilde{v})$$

The definition of the transition relation $\rightarrow$ is provided in Table 5 and relies on the rules Comp, FComp, Sync, ComL and ComR.

Rules Comp and FComp depends on relation $\mapsto$ and are used to lift the effect of local behaviour to the system level. The former states that the relations $\mapsto$ and $\rightarrow$ coincide when performing either an input or an output actions, while rule FComp states that a component $\Gamma :_I P$ can discard a message and remain unchanged. However, we would like to stress that the system level label of FComp coincides with that of Comp in case of input actions, which means that externally it cannot be percieved whether a message has been accepted or discarded.

Rule Sync states that two parallel components $C_1$ and $C_2$ can receive the same message.

$$\frac{\Gamma:_I P \xmapsto{\lambda} \Gamma':_I P'}{\Gamma:_I P \xrightarrow{\lambda} \Gamma':_I P'} \text{Comp} \qquad \frac{\Gamma:_I P \xmapsto{\widetilde{\Gamma' \triangleright \Pi'(\tilde{v})}} \Gamma:_I P}{\Gamma:_I P \xrightarrow{\Gamma' \triangleright \Pi'(\tilde{v})} \Gamma:_I P} \text{FComp}$$

$$\frac{C_1 \xrightarrow{\Gamma \triangleright \Pi(\tilde{v})} C_1' \quad C_2 \xrightarrow{\Gamma \triangleright \Pi(\tilde{v})} C_2'}{C_1 \parallel C_2 \xrightarrow{\Gamma \triangleright \Pi(\tilde{v})} C_1' \parallel C_2'} \text{Sync}$$

$$\frac{C_1 \xrightarrow{\Gamma \triangleright \overline{\Pi}(\tilde{v})} C_1' \quad C_2 \xrightarrow{\Gamma \triangleright \Pi(\tilde{v})} C_2'}{C_1 \parallel C_2 \xrightarrow{\Gamma \triangleright \overline{\Pi}(\tilde{v})} C_1' \parallel C_2'} \text{ComL} \qquad \frac{C_1 \xrightarrow{\Gamma \triangleright \Pi(\tilde{v})} C_1' \quad C_2 \xrightarrow{\Gamma \triangleright \overline{\Pi}(\tilde{v})} C_2'}{C_1 \parallel C_2 \xrightarrow{\Gamma \triangleright \overline{\Pi}(\tilde{v})} C_1' \parallel C_2'} \text{ComR}$$

Table 5. Operational Semantics of Systems

Finally, rule ComL (and its symmetric variant ComR) governs communication among two parallel components $C_1$ and $C_2$: If $C_1$ sends a message then ($C_2$ can receive it by applying rule Comp). However, $C_2$ has also the possibility of discarding the message by applying rule FComp.

## 2.3 Expressiveness of AbC Calculus

In this section, we would like to provide some evidences of the expressive power of the *AbC* calculus by showing how different communication models and interaction patterns can be easily modelled in *AbC*. Indeed, we think that attribute-based communication can be used as a unifying framework to encompass a number of communication models. Further details regarding the actual implementation of the material, presented in this section, can be found in the Webpage of $Ab^a CuS^3$: a Java API for the *AbC* calculus.

*Encoding channel-based interaction.* We show how one-to-many *channel-based interaction* can be encoded in the *AbC* calculus. It may seem tempting to model a channel name as an attribute in *AbC*, however it turns out not to be the case. The reason is that in channel-based communication, a channel, where the exchange happens, is instantly enabled at the time of interaction and is disabled afterwards. This feature is not present in *AbC* since attributes are persistent in the attribute environment and cannot be disabled at any time (i.e., attribute values are always available to be checked against sender predicates). However, this is not a problem because we can exploit the fact that the receiving predicates in *AbC* can check the values in the received message. The key idea is to use structured messages to select communication partners where the name of the channel is rendered as the first element in the message; receivers only accept messages with attached channels that match their receiving channels. Actually, attributes do not play any role in such interaction so we assume components with empty environments and interfaces i.e., $\emptyset :_\emptyset P$. In what follows, we use $[P]$ to denote $\emptyset :_\emptyset P$. Thus a pair of processes, one willing to receive on channel $a$ and the other willing to send on the same channel, can be modeled as follows:

$$[(x = a)(x, y).P] \parallel [(a, msg)@(\text{tt}).Q]$$

---
[3] http://lazkany.github.io/AbC/

To show the feasibility of encoding broadcast channel-based calculi into $AbC$, we have encoded the $b\pi$-calculus [Ene and Muntean 2001] into $AbC$. The $b\pi$-calculus has been chosen because it uses broadcast instead of binary communication as a basic primitive for interaction which makes it a sort of variant of value-passing CBS [Prasad 1991]. Furthermore, channels in $b\pi$-calculus can be communicated like in the point-to-point $\pi$-calculus [Milner et al. 1992] which is considered as one of the richest paradigms introduced for concurrency so far.

Based on a separation result presented in [Ene and Muntean 1999], it has been proven that $b\pi$-calculus and $\pi$-calculus are incomparable in the sense that there does not exist any uniform, parallel-preserving translation from $b\pi$-calculus into $\pi$-calculus up to any "reasonable" equivalence. On the other hand, in $\pi$-calculus a process can non-deterministically choose the communication partner while in $b\pi$-calculus it cannot. Proving the existence of a uniform and parallel-preserving encoding of $b\pi$-calculus into $AbC$ up to some reasonable equivalence ensures at least the same separation results between $AbC$ and $\pi$-calculus.

The full encoding, the formal definition which specifies what properties are preserved by this encoding, and a proof for its correctness up to a specific behavioural equivalence can be found in [Alrahman et al. 2017a].

*Encoding interaction patterns.* Also *group-based* [Agha and Callsen 1993; Chockler et al. 2001; Holbrook and Cheriton 1999] and *publish/subscribe-based* [Bass and Nguyen 2002; Eugster et al. 2003] interaction patterns can be naturally rendered in $AbC$. In the group-based model, when an agent wants to send a message to all elements of a group, it attaches the name or a reference to the group in the message and the message is propagated using this reference. In the publish/subscribe model, there are two types of agents: publishers and subscribers and there is an exchange server that mediates their interactions. For instance, in topic-based publish/subscribe models [Eugster et al. 2003], publishers produce messages tagged with topics and send them to the exchange server which is responsible for filtering and forwarding these messages to interested subscribers. Subscribers simply register their interests to the exchange server and receive messages according to their interests. Since these interaction patterns do not have formal descriptions, we proceed by relying on examples.

We start with group-based interaction patterns and show that when using a group name as an attribute in $AbC$, the constructs for joining or leaving a group can be modelled as attribute updates, like in the following example, where we assume that initially we have $\Gamma_1(group) = b$, $\Gamma_2(group) = a$, and $\Gamma_7(group) = c$:

$$\Gamma_1:_{\{group\}}(msg)@(group = a).0 \parallel$$
$$\Gamma_2:_{\{group\}}(group = b)(x).0 \mid \mathsf{set}(\mathsf{this}.group, c)0 \parallel \ldots$$
$$\parallel \Gamma_7:_{\{group\}}(group = b)(x) \mid \mathsf{set}(\mathsf{this}.group, a)0$$

Component 1 wants to send the message "$msg$" to group "$a$". Only Component 2 is allowed to receive it as it is the only member of group "$a$". Component 2 can leave group "$a$" and join "$c$" by performing an attribute update as reported on the right hand side of the interleaving operator | . On the other hand, if Component 7 joined group "$a$" before "$msg$" is emitted then both of Component 2 and Component 7 will receive the message.

It is worth mentioning that a possible encoding of group communication into $b\pi$-calculus has been introduced in [Ene and Muntean 2001]. The encoding is relatively complicated and does not guarantee the causal order of message reception. "Locality" is neither a first class construct in $b\pi$-calculus nor in $AbC$. However, "locality" (the group name, in this case) can be naturally modeled as an attribute in $AbC$; in $b\pi$-calculus, more effort is needed.

Publish/subscribe interaction patterns can be considered as special cases of the attribute-based ones. For instance, a natural modeling of the topic-based publish/subscribe model [Eugster et al. 2003] into $AbC$ can be obtained by allowing publishers to broadcast messages with true predicates (i.e., satisfied by all subscribers) and requiring subscribers to check compatibility of the exposed publishers attributes with their subscriptions, like in the example below:

$$\Gamma_1:_{\{topic\}}(msg)@(\text{tt}).0 \parallel \Gamma_2:_{\{subscription\}}(topic = \text{this}.subscription)(x).P \parallel$$
$$\ldots \parallel \Gamma_n:_{\{subscription\}}(topic = \text{this}.subscription)(x).Q$$

The publisher broadcasts the message "*msg*" tagged with a specific topic for all subscribers (predicate "tt" is satisfied by all); subscribers receive the message if the topic matches their subscription.

In the next section, we present a case study inspired by an industrial application of CAS. We focus on the role of the communication primitives and of the external environment in determining the communication between interacting components. Moreover, we use the case study to motivate the new communication primitives of *AbC* and to show that attribute-based communication is appropriate for handling interactions in CAS.

## 3 STABLE ALLOCATION IN CONTENT DELIVERY NETWORKS

This case study is based on the distributed stable allocation algorithm adopted by Akamai's Content Delivery Network (CDN) [Maggs and Sitaraman 2015]. Akamai's CDN is one of the largest distributed systems in the world. It has currently over 170,000 servers located in over 1300 networks in 102 countries and serves 15-30 % of all Web traffic. To avoid dealing with billions of clients individually, Akamai divides the clients of the global internet into groups, called *map units* each having a specific demand, based on their locations and traffic types. Also content servers are grouped into clusters, and each cluster is rated according to its capacity, latency, etc. Map units prefer highly rated clusters while clusters prefer low demand map units. The goal of global load balancing is to assign map units to clusters such that preferences are accounted for and constraints are met.

The allocation algorithm in [Maggs and Sitaraman 2015] is a slight variant of the original Stable Marriage Problem (SMP), reported in [Gale and Shapley 1962]. The goal of the original algorithm is to find a stable matching between two equally sized sets of elements given an ordering of preferences of each element of the two sets. Each element in one set has to be paired to an element in the opposite set in such a way that there are no two elements of different pairs which both would rather have each other than their current partners. When there are no such pairs of elements, the set of pairs is deemed stable. A natural and straightforward *AbC* implementation of the original stable marriage problem can be found in [Alrahman et al. 2016b].

The variant considered in [Maggs and Sitaraman 2015] allows (1) more map units to be assigned to a single cluster and (2) map units to rank only the top dozen, or so, clusters that are likely to provide good performance. The first feature is a typical generalisation [Iwama and Miyazaki 2008] of the original SMP, while the second is a mere simplification of the problem. Implementing these features in *AbC* does not pose any challenge, but it would make the example more verbose. Actually, our implementation of the original problem in [Alrahman et al. 2016b] needs only to be extended to consider an extra attribute, named capacity, necessary for determining when a cluster should stop engaging with more map units. Moreover, the map units assigned to a cluster should be ordered according to their demands, so that a dissolve message goes first to the most demanding map units when necessary. Furthermore, these features do not add much to the original SMP; they still require map units and clusters to have predefined lists of preferences such that only ranked elements can participate in the algorithm. Obviously, this implies that one cannot take advantage of dynamic creation of new clusters.

In this article, we consider a more interesting variant of SMP [Gale and Shapley 1962] that is better suited for the dynamicity of CDN. In this variant, the arrival of new clusters is considered, it is not required that elements know each other, and no predefined preference list is assumed. Notice that, in these settings, point-to-point interaction is not possible because elements are not aware of each other and the choice of implicit multicast is crucial. Indeed, in our variant, elements express their interests in potential partners by relying on their attributes rather than on their identities. In essence, an element of one set communicates with elements of the opposite set using predicates. Two parties that agree on some predicates form a pair, otherwise they keep looking for better partners. A pair splits only if one of its elements can find a better partner willing to accept its offer. In this way, preferences are represented as predicates over the values of some attributes of the interacting partners.

In this scenario, we consider the values of attributes *demand*, for a map unit, and *rating*, for a cluster, as a means to derive the interaction. For simplicity, these attributes can take two different values: high ($H$) and low ($L$). An element in the system can be either a *Unit* or a *Cluster*. Units start the protocol by communicating with clusters in the quest of finding an element that satisfies their highest expectations. If no cluster accepts the offer, a unit lowers its expectations and proposes again until a partner is found. Clusters are always willing to accept proposals from any unit that enhances their levels of satisfaction. In case of a new arrival, some pairs of elements might dissolve if the new arrival enhances their levels of satisfaction. This means that not all pairs in the system are required to split on new arrivals; only those interested will do so. The system level goal (the emergent behaviour) is to construct a set of stable pairs from elements of different types by combining the behaviour of individual elements in the system through message passing. Mathematically speaking, the problem consists of computing a function at the system level by combining individual element behaviours, without relying on a centralised control. Notice that since map units initiate the interaction, the solution is a "map-unit-optimal", as proved in [Gale and Shapley 1962], which is a property that fits with the CDN's goal of maximising performance for clients.

Allowing new arrivals is crucial to guarantee scalability and open-endedness while communicating based on predicates rather than on identities or ranks is crucial to deal with anonymity. The actual implementation of this algorithm alongside with experiment results can be found in the Webpage of GoAt[4]: a Go API for the *AbC* calculus. There we also provide a short tutorial to provide some intuition about how to use this API for programming.

The system in our attribute-based scenario can be modelled in *AbC* as the parallel composition of existing units and clusters (i.e., $Unit_i \parallel \ldots \parallel Unit_n \parallel Cluster_i \parallel \ldots \parallel Cluster_n$). Notice that units and clusters interact in a distributed setting without any centralised control. Each element is represented as an *AbC* component. A unit, $Unit_i$, has the form $\Gamma_i :_I P$ where $\Gamma_i$ represents its attribute environment, $I$ represents its interface where $I = \{demand, id_i\}$, and the process $P$ represents its behaviour. A cluster, $Cluster_i$, has the form $\Gamma_r :_{I'} Q$ where $\Gamma_r$ represents its attribute environment, $I'$ represents its interface where $I' = \{rating, id_r\}$, and the process $Q$ represents its behaviour.

In addition to the attributes *demand* and *rating*, mentioned above, the attribute environments of units and clusters contain the following attributes:

**partner**: current partner's identity; in case they are not engaged, the value is $-1$;
**exPartner**: previous partner;
**id$_i$** and **id$_r$**: the identity of units and clusters, respectively;
**ref**: current preference, 0 for high rating and 1 for low rating, initially $ref = 0$;

---

[4]https://github.com/giulio-garbi/goat

- **success**: a boolean attribute which is set to true when an accept message from a cluster is received, initially $success = \text{ff}$;
- **arrival**: a boolean attribute which is set to true when an arrival message is received, initially $arrival = \text{ff}$;
- **ack**: a boolean attribute which is set to false when a negative acknowledgment message is received, initially $ack = \text{tt}$;
- **dissolve**: a boolean attribute which is set to true when a dissolve message is received, initially $dissolve = \text{ff}$;
- **rank**: an integer flag used to rank the current partner: 0 for high and 1 for low, initially $rank = 2$;
- **bof**: an integer flag used to rank the new arrival: 0 for high and 1 for low. Initially $bof = 2$;
- **lock**: an integer flag used to implement a lock within a single component, initially $lock = 0$;
- **timer**: a counter implementing a timer of a single component. Initially $timer > time\_out$ where $time\_out$ is a constant number representing the number of steps a process can take before specific event can occur.

The behaviour of a unit component is specified by the process $P$ which is the parallel composition of the processes $I$, $T$, $M$, and $N$. Process $I$ defines a proposal process, process $T$ defines a timer process, process $M$ defines a message handler process, and process $N$ defines a negative acknowledgment process. The behaviour of the proposal process, $I$, is defined below:

$$I \triangleq \langle ref = 0 \rangle (\text{``propose''}, \text{this}.demand, \text{this}.id_i, 0)@\Pi_h.[timer := 0, dissolve := \text{ff}]I'$$
$$+$$
$$\langle ref = 1 \rangle (\text{``propose''}, \text{this}.demand, \text{this}.id_i, 1)@\Pi_l.[timer := 0, dissolve := \text{ff}]I'$$

Process $I$ sends a proposal message to all components that either satisfy predicate $\Pi_h$ or predicate $\Pi_l$, depending on the current value of the $ref$ attribute. The predicate $\Pi_h$ represents high expectations where $\Pi_h = (rating = \text{``H''})$ while the predicate $\Pi_l$ represents low expectations where $\Pi_l = (rating = \text{``L''})$. Notice that the branches of process $I$ encode the preferences of a unit and the selection of any of them depends on the run-time value of $ref$. These branches can be thought of as context-dependent behavioural variations in Context-Oriented Programming [Hirschfeld et al. 2008]. Since the initial value of $ref$ is 0, the process proceeds with the first branch. The proposal message contains a proposal label, "*propose*", the values of attributes $demand$, $id_i$, and $ref$ of the unit respectively. The sent value of $ref$ will be used later to decide if an accept message from a cluster is stale (i.e., the received value of $ref$ is different from the current value of $ref$). By sending a proposal message, the timer and the dissolve attributes are reset. After this step, process $I$ evolves to $I'$ which is a proposal handler process. Resetting the timer attribute will decide how long process $I$ should wait before proposing again. The timer process is reported below:

$$T \triangleq \langle timer < time\_out \rangle ()@\text{ff}.[timer := timer + 1]T$$
$$+$$
$$\langle timer = time\_out \rangle ()@\text{ff}.[timer := timer + 1, ref := (ref + 1)\%2, success := \text{ff}]T$$

$T$ keeps increasing the value of the timer autonomously with silent moves (i.e., ()@ff) as long as its value is less than the constant $time\_out$. By assuming fairness and a reasonably large value of $time\_out$, we can guarantee that an accept message from a cluster, that satisfies the sent proposal message (if there is any), is received before a timeout. If a timeout is reached, the timer process lowers the preference of the unit and resets the value of the success attribute to false. As we have seen so far, the code of processes is infrastructure-agnostic, i.e., it does not contain addresses or channel names. It is completely data-centric and relies on the run-time characteristics of the interacting partners.

In the proposal handler process, reported below, we can understand the role of the awareness construct $\langle \Pi \rangle$ as an environmental parameter used to influence the behaviour of a unit at run-time.

$$I' \triangleq \langle lock = 0 \wedge \neg success \wedge timer > time\_out \rangle I$$
$$+$$
$$\langle lock = 0 \wedge dissolve \rangle I$$
$$+$$
$$\langle lock = 0 \wedge arrival \wedge (bof \leq rank - 1) \rangle$$
$$(\text{``}dissolve\text{''})@(id_r = \text{this}.partner).$$
$$[arrival := \text{ff}, success := \text{ff}, ref := bof, bof := 2, rank := 2,$$
$$exPartner := partner, partner := -1]I$$

The process blocks executing until one of three events occurs. If no lock is acquired ($lock = 0$), no accept message from a cluster is received ($\neg success$), and a timeout occurs ($timer > time\_out$), process $I'$ calls the proposal process $I$ again by considering the new value of $ref$ modified by the timer process. If no lock is acquired and a dissolve message from the current partner is received ($dissolve = true$), process $I'$ calls the proposal process $I$ again. Finally, if no lock is acquired and an arrival message is received ($arrival = true$), and the rank of the new arrival is better than the rank of the current partner ($bof \leq rank - 1$), the process sends a dissolve message that contains a *dissolve* label to its partner, sets the value of attribute *exPartner* to the value of attribute *partner* and the value of attribute $ref$ to the value of attribute $bof$, and resets the values of attributes *arrival*, *success*, *bof*, *rank*, and *partner* to their initial values. Process $I'$ calls the proposal process $I$ again.

The negative acknowledgment process is reported below:

$$N \triangleq (x = \text{``}accept\text{''} \wedge ((z \neq ref) \vee success))(x, y, z).$$
$$(\text{``}ack\text{''}, -1)@(id_r = y).0 \mid N$$

$N$ ensures that after a successful reception of a first accept message from a cluster (i.e., $success = \text{tt}$) all other accept messages to this unit are discarded (i.e., it sends an acknowledgement message with id=-1 which is interpreted by a cluster as a negative acknowledgement). Notice that the process replicates itself every time an accept message is received to be always able to catch all accept messages; the condition ($z \neq ref$) is crucial to discard stale messages.

The message handler process is reported below:

$$M \triangleq \langle \neg success \rangle (x = \text{``}accept\text{''} \wedge z = ref)(x, y, z).$$
$$[lock := 1, success := \text{tt}, rank := ref, exPartner := partner, timer := time\_out + 1]$$
$$(\text{``}ack\text{''}, y)@\text{tt}.[lock := 0, parnter := y]M$$
$$+$$
$$(x = \text{``}dissolve\text{''} \wedge id_i = partner)(x).$$
$$[dissolve := \text{tt}, success := \text{ff}, ref := 0, rank := 2, partner := -1]M$$
$$+$$
$$\langle partner \neq -1 \rangle (x = \text{``}arrived\text{''} \wedge rating = \text{``}H\text{''})(x).[arrival := \text{tt}, bof := 0]M$$

$$+$$
$$\langle bof \neq 0 \wedge partner \neq -1 \rangle (x = \text{``arrived''} \wedge rating = \text{``L''})(x).[arrival := \text{tt}, bof := 1]M$$

$M$ can respond to one of three events: if no accept message is received yet (i.e., $\neg success$) and a new one arrives, the process receives the accept message only if the message is not stale ($z = ref$), acquires a lock ($lock := 1$), sets the value of attribute $success$ to true, the value of attribute $rank$ to the value of attribute $ref$, the value of attribute $exPartner$ to the value of attribute $partner$ and turns off the timer. The process proceeds by sending an acknowledgement message to the cluster, releasing the lock, setting the value of attribute $partner$ to the received cluster identity, and then the process continues as $M$. Notice that the lock is important to ensure that the handler process $I'$ does not proceed before all required attributes are assigned the right values. In this case, the first branch of process $M$ is executed atomically with respect to the co-located processes in the unit component.

If a dissolve message from the current partner is received, the process sets the value of attribute $dissolve$ to true, resets the values of attributes $success$, $ref$, $rank$, and $partner$ to their initial values and continues as $M$. If an arrival message is received, the arrival attribute is set to true and the value of attribute $bof$ is set to 0 if the attached attribute value is high $rating = \text{``H''}$ otherwise the value of attribute $bof$ is set to 1 and the process continues as $M$. Notice that arrivals with low value of attribute $rating$ cannot override the ones with high values. This is guaranteed by the condition $bof \neq 0$ on the last branch. Also an arrival message can have an effect on the behavior of a unit only if the unit is already engaged otherwise the arrival message is just discarded which is ensured by the condition $partner \neq -1$ in the last two branches of process $M$.

The coordination among processes running in a single unit is made possible by relying on shared attributes, awareness constructs, and attribute updates to implement proper lock mechanisms.

The behaviour of a cluster component is specified by the process $Q$ which is the parallel composition of the processes $A$, $R$, and $D$. Process $A$ defines an arrival process, process $R$ defines a proposal reception process, and process $D$ defines a dissolve handler process. The arrival process, $A$, is defined below:

$$A \triangleq (\text{``arrived''})@\text{tt}.0$$

This is the first process that takes a step when a cluster component wants to be involved in the allocation procedure. It simply sends an arrival message to all components in the system and then terminates. The arrival message contains an arrival label, "$arrived$".

The proposal reception handler is reported below:

$$R \triangleq (x = \text{``propose''})(x,\ y,\ z,\ n).R' \mid R$$

The process simply receives a proposal message and evolves to the reception handler process $R'$. Notice that process $R$ replicates itself every time a message is received to ensure that no proposal is lost. The reception handler process, $R'$ reported below, checks if no lock is acquired and the rank of the communicated value of attribute $demand$ of the unit is less than the rank of the current partner ($rank(y) < rank$), the process sends an accept message to the unit where the proposal message came from, addressing it by its identity and acquires a lock, otherwise the process $R'$ terminates. The function $rank(y)$ takes the value of $demand$ as a parameter and returns 1 if $y = \text{``H''}$ and 0 otherwise. The accept message contains an acceptance label, "$accept$", the identity of the current cluster, and the reference of the proposal message. The process then waits to receive either acknowledgement message or a negative one from the unit. If an acknowledgement is received, the value of attribute partner is set to the received identity of the unit, the rank is set to the value returned by $rank(y)$, the lock is released and process $R'$ terminates. Notice that a dissolve message is sent to the current partner before releasing the lock in case a cluster is already engaged. If a negative acknowledgement is received then the lock is released. The lock is needed to ensure that concurrent proposals are handled sequentially which is important to guarantee a consistent state of the attribute environment.

$$
\begin{aligned}
R' \triangleq \ &\langle lock = 0 \rangle ( \\
&\textbf{if } (rank(y) < rank) \textbf{ then } \{ \\
&\quad (\text{``accept''}, id_r, n)@(id_i = z).[lock := 1]( \\
&\quad \textbf{if } (partner \neq -1) \textbf{ then } \{ \\
&\quad\quad (e = \text{``ack''} \land f = id_r \land id_i = z)(e, f). \\
&\quad\quad [exPartner := partner, \ partner := z, \ rank := rank(y)] \\
&\quad\quad (\text{``dissolve''})@(id_i = \text{this}.exPartner).[lock := 0]0 \\
&\quad \} \textbf{ else } \{ \\
&\quad\quad (e = \text{``ack''} \land f = id_r \land id_i = z)(e, f). \\
&\quad\quad [partner := z, \ rank := rank(y), \ lock := 0]0 \\
&\quad \} \\
&\quad + \\
&\quad (e = \text{``ack''} \land f \neq id_r \land id_i = z)(e, f).[lock := 0]0) \\
&\} \textbf{ else } 0)
\end{aligned}
$$

The dissolve handler process is reported below:

$$
\begin{aligned}
D \triangleq \ &(x = \text{``dissolve''} \land id_i = partner)(x). \\
&[rank := 2, \ exPartner := partner, \ partner := -1]D
\end{aligned}
$$

When a dissolve message from a cluster's partner is received, the value of attribute exPartner is set to the value of attribute partner, the values of attributes rank and partner are reset to their initial values, and the process calls itself.

REMARK 3.1. *In [Gale and Shapley 1962], the authors showed that their algorithm terminates with a matching that is stable after no more than $n^2$ proposals, where n is the number of proposing elements, i.e., the algorithm has $O(n^2)$ worst-case complexity. In our variant, it should be clear that the worst case complexity is also $O(n^2)$ even after relaxing the assumptions of the original algorithm, i.e., no predefined preference lists and components are not aware of the existence of each other, so point-to-point communication is not possible. Interestingly, the complexity is still quadratic even if we consider a blind broadcast mechanism where proposals are sent to all components in the system except for the sender unit. In this way, for n-units, n-clusters, and a constant L related to the preferences of a unit where L is computed based on the number of branches in the proposal process I, we have that each unit can send at most $L * (2n − 1)$ proposals.*

Below we first provide a simple interaction fragment to show how the semantics rules are applied in our stable allocation (Example 3.1). We then show a possible and complete execution of the protocol by considering two pairs of components (Example 3.2).

*Example 3.1 (Interaction fragment).* A unit, say $I_1$ with $ref = 1$, $demand = $ "H", and $id_i = 1$, can send a proposal and generate the following transition by applying rule Comp:

$$\Gamma_i:_{\{demand, id_i\}} I \mid T \mid M \mid N \xrightarrow{\{demand=\text{``H''}, id_i=1\} \triangleright \overline{\Pi_l}(\text{``propose''}, \text{``H''}, 1, 1)} $$
$$\Gamma_i[timer \mapsto 0, dissolve \mapsto \text{ff}]:_{\{demand, id_i\}} I' \mid T \mid M \mid N$$

Any cluster, say $r_1$, can receive the message by applying rule Comp and generate this transition:

$$\Gamma_r:_{\{rating, id_r\}} R \mid A \mid D \xrightarrow{\{demand=\text{``H''}, id_i=1\} \triangleright \Pi_l(\text{``propose''}, \text{``H''}, 1, 1)}$$
$$\Gamma_r:_{\{rating, id_r\}} R'[\text{``propose''}/x, \text{``H''}/y, 1/z, 1/n] \mid R \mid A \mid D$$

Now the overall system evolves by applying rule ComL as follows:

$$S \xrightarrow{\{demand=\text{``H''}, id_i=1\} \triangleright \overline{\Pi_l}(\text{``propose''}, \text{``H''}, 1, 1)} \Gamma_i[timer \mapsto 0, dissolve \mapsto \text{ff}]:_{\{demand, id_i\}} I' \mid T \mid M \mid N$$
$$\parallel \dots \parallel \Gamma_r:_{\{rating, id_r\}} R'[\text{``propose''}/x, \text{``H''}/y, 1/z, 1/n] \mid R \mid A \mid D$$

□

*Example 3.2.* Let us consider two map units $M_0$ and $M_1$ and two clusters $C_0$ and $C_1$ with attribute environments defined as follows: $\Gamma_{m_0} = \{demand = \text{``H''}, id_i = m_0\}$, $\Gamma_{m_1} = \{demand = \text{``L''}, id_i = m_1\}$, $\Gamma_{c_0} = \{rating = \text{``H''}, id_r = c_0\}$, and $\Gamma_{c_1} = \{rating = \text{``L''}, id_r = c_1\}$. Below we show the execution of the stable allocation protocol, presented above, until a stable matching is reached.

- $C_0$ broadcasts an arrival message through process $A$ by applying Comp.
- $C_1$ broadcasts an arrival message through process $A$ by applying Comp. Because no unit is engaged yet, all arrival messages are discarded.
- $M_0$ proposes for a cluster with a high rating by applying Comp and waits for a response.
- $C_0$ receives the proposal because its rating is high by applying Comp.
- $M_1$ proposes for a cluster with a high rating by applying Comp and waits for a response.
- $C_0$ receives the proposal because its rating is high by applying Comp.
- $C_0$ handles the first proposal by process $R'$. It acquires a lock and sends an accept message to $M_0$ by applying rule Comp. Notice that the second proposal cannot be handled before the lock is released.
- $M_0$ receives the message by process $M$, acquires a lock and sets the attributes success, rank and exPartner appropriately by rule Comp.
- $M_0$ sends an acknowledgement message to all possible components and includes the identity of $C_0$ in the message. It sets its partner to $C_0$ and releases the lock.
- Clearly, all components are potential receivers for this message, however, only $C_0$ receives it because it is the only component waiting an acknowledgement that contains its identity. By doing so, $C_0$ sets its partner to $M_0$, its rank to 1 (because $rank(\text{``H''}) = 1$) and releases the lock.
- $C_0$ handles the second proposal by process $R'$ again. It acquires a lock and sends an accept message to $M_1$ because $rank(\text{``L''}) = 0 < 1$, by applying rule Comp and waits for an acknowledgement.
- $M_1$ is the only component that can receive the accept message because it satisfies the sending predicate by applying rule Comp. It receives the message by process $M$, acquires a lock and sets the attributes success, rank and exPartner appropriately by rule Comp.

- $M_1$ sends an acknowledgement message to all possible components and includes the identity of $C_0$ in the message. It sets its partner to $C_0$ and releases the lock.
- $C_0$ receives the acknowledgement, sets its exPartner to current partner, its partner to $M_1$ and its rank to 0 by rule Comp.
- $C_0$ sends a dissolve message to its exPartner (i.e., $M_0$) and releases the lock by rule Comp.
- $M_0$ receives the dissolve message by the second branch of process $M$ which sets the dissolve attribute to true and resets the attributes *success*, *ref*, *rank*, and *partner* to their initial values. By doing so, the second branch of process $I'$ will be enabled and $M_0$ is ready to propose again.
- $M_0$ proposes again for a cluster with a high rating by applying Comp and waits for a response.
- $C_0$ receives the proposal because its rating is high by applying Comp.
- $C_0$ drops the proposal by the last branch of process $R'$, because it is satisfied with its current partner, i.e., its rank<1.
- No component can actually respond to the proposal. $M_0$ keeps waiting until a timeout occurs, i.e., the second branch of process $T$ is enabled. The timer is turned off, the ref attribute value is set to 1 and the success attribute is turned off. $M_0$ can now propose for a cluster with a low rating. In this case, only $C_1$ accepts the proposal in a similar way as described before. Notice that $(M_0, C_1)$ and $(M_1, C_0)$ are actually stable pairs.

## 4 DISCUSSION

In this section, we first explain how *AbC* has helped us in dealing with the collective adaptive features that were evident in the previous case studies, finally we comment on other features of *AbC* code, namely *Compositionality* and *Code Extensibility*.

*Distribution* is naturally obtained because an *AbC* system has no centralised control; and it is represented as the parallel composition of independent components that can mutually influence each other only through message-exchange. For instance in Section 3, the overall system is defined as the parallel composition of existing units and clusters.

*Awareness* is supported by the attribute environment that plays a crucial role in orchestrating the behaviour of *AbC* components. It makes components aware of their own status and provides partial views of the surrounding environment. Components behave differently under different environmental contexts. This is possible because the behaviour of *AbC* processes is parametric with respect to the run-time attribute values of the component in which they are executed. The awareness construct, $\langle \Pi \rangle$, is used as an environmental parameter to influence the behaviour of *AbC* components at run-time. For instance, in Section 3, the behaviour of the proposal handler process, $I'$, is totally dependent on the values of the attributes *lock*, *success*, *timer*, *dissolve*, *arrival*, *bof*, and *rank*. In case the value of just one of these attributes changes, the process will change its behaviour accordingly.

*Adaptivity* is obtained by means of the interaction predicates (both for sending and receiving) of *AbC* components that can be parametrised with their local attribute values; any run-time changes of these values might influence/change the possible set of interaction partners. Notice that the target of the dissolve message, in the third branch of the proposal handler process $I'$ in Section 3, depends on the identity of the current partner (this.*partner*).

*Interdependence* among co-located processes can be obtained by modifying, with the attribute update construct, the attribute environment shared among processes within a single component. A branch of process $I'$, in Section 3, is mainly selected depending on the attribute updates performed by the timer $T$ and the message handler $M$.

*Collaboration* is obtained by combining individual component behaviours, through message exchange, to achieve a global goal for which a single component would not be sufficient. For instance, in Section 3, the goal was to construct a set of stable pairs without any centralised control. The goal was achieved by allowing different components to collaborate through message passing; each component contributed, and the combined behaviour of all components was necessary, to reach the overall system goal.

*Anonymity* is obtained by allowing the interaction primitives (both send and receive) to rely on predicates over the run-time attribute values of the interacting partners rather than on specific names or addresses. Thus the qualification needed to receive a message no longer depends on available channels or addresses, but on the values of dynamic attributes. For instance, in Section 3, the process $I$ sends a proposal to a group of clusters whose attributes satisfy a specific predicate (i.e., $\Pi_h$ or $\Pi_l$). There is no prior agreement between the units and the other clusters and the set of candidate receivers is specified at the time of interaction in the sense that any change in the value of attribute $ref$ changes the set of targeted components.

*Scalability and Open-endedness* are guaranteed by the adoption of multiparty communication instead of the binary one. Actually *AbC* supports an implicit multicast communication in the sense that the multicast group is not specified in advance, but rather it is specified at the time of interaction by means of the available set of receivers with attributes satisfying the sender predicate. The non-blocking nature of the *AbC* multicast, alongside with anonymity of interaction, breaks the synchronisation dependencies between senders and receivers and components can join or leave a system without disrupting its overall behaviour. For instance, new clusters, as defined by process $A$, Section 3, can join the allocation procedure at run-time without causing any disruption to the overall system behaviour. Actually, the third branch of process $I'$ of a unit is used as an adaptation mechanism to handle new arrival of clusters. Notice that the condition $bof \leq rank - 1$ of that branch ensures that a unit responds only to new clusters that enhance its satisfaction. This means that new arrivals affect only specific components in the system.

It is worth mentioning that the initial behaviour of all components with the same type (e.g., Unit or Cluster in Section 3) is exactly the same. However, since this behaviour is parametrised with respect to the run-time attribute values of each component, components might exhibit different behaviours. This means that the context where a component is executing has a great impact on its behaviour, in the sense that it can disable or enable specific behaviours depending on the environmental changes. In some sense, the behaviour of a component evolves based on contextual conditions. Components do not need to have complex behaviour to achieve adaptation at system level. Complex behaviour can be achieved by combining the local behaviour of individual components.

We would also like to stress that *AbC* code naturally supports compositionality and extensibility.

*Compositionality*: a component is the basic building block of *AbC* programs and any program can be broken down into a set of individual components which can only interact by exchanging messages. This simplifies verification because components can be analysed individually and their external behaviour can be assessed by their observable communication capabilities. Actually, we can abstract from the internal behaviour of an *AbC* component and only consider its observable behaviour when it interacts with other components. Thus, a component can be treated as a blackbox that acts and reacts to its environment. We refer interested readers to our theoretical results in [Alrahman et al. 2016a, 2017a].

*Code Extensibility*: we would like to stress that *AbC* code can be easily extended in the sense that new alternative behaviours can be added and removed by modifying the attribute environments and the interfaces of components. Also component behaviours can be adapted without changing the internal structure of their running processes. For instance in Section 3, a unit can reverse the order of its preferences by directly modifying the required value of attribute *ref* in the awareness constructs and the following output actions of process *I*. Also if we add a public attribute to a cluster, say *location*, the code of a unit can be easily adapted to consider new possible preferences. One way is to add another alternative/branch to process *I* that consider the location of a cluster. Of course some attribute values have to be adjusted accordingly. As mentioned before, these branches can be thought of as context-dependent behavioural variations in Context-Oriented Programming [Hirschfeld et al. 2008].

## 5 RELATED WORK

In this section, we touch on related works concerning languages and calculi with primitives that model either collective interaction, or multiparty interaction. We also report on other approaches to interaction in distributed systems and show how they relate to *AbC*.

Some of the well-known approaches include: channel-based models [Hoare 1978; Milner 1980; Milner et al. 1992], group-based models [Agha and Callsen 1993; Chockler et al. 2001; Holbrook and Cheriton 1999], and publish/subscribe models [Bass and Nguyen 2002; Eugster et al. 2003]. Below we briefly report the main features and limitations of such approaches.

*Channel-based* models rely on explicit naming for establishing communication links in the sense that communicating partners have to agree on channels or names to exchange messages. This implies that communicating partners are not anonymous to each other. Actually, communicating partners have static interfaces that determine the target of communication e.g., binary communication CCS [Milner 1980], multiway synchronisation CSP [Hoare 1978], and broadcast communication CBS [Prasad 1995]. In *AbC* the interacting partners are anonymous to each other; they do not need to agree on channels or names to interact, but rather to rely on the satisfaction of predicates over the attributes they expose. This makes our calculus more suitable for modeling scalable distributed systems as anonymity is a key factor for scalability. The $\pi$-calculus [Milner et al. 1992] was developed as a way to mitigate this problem by allowing the exchange of channel names and thus providing dynamic interfaces and additional flexibility. However, the dynamicity of interfaces is still limited because even if generic input or output actions are allowed, they are disabled until they are instantiated with specific channel names. This means that a process can engage in a communication only when its actions are enabled. The broadcast $b\pi$-calculus [Ene and Muntean 1999, 2001] is based on CBS and the $\pi$-calculus in that it extends the former with a channel-passing mechanism. Furthermore, $\pi$-calculus and most process calculi rely on synchronous communication that creates dependencies between senders and receivers and affects the overall scalability of the system. *AbC* actions are always enabled with respect to the current attribute values of the component where they are executing. When these values change, the interaction predicates change seamlessly and become available for other communication partners.

In *group-based* models, like the one used for IP multicast [Holbrook and Cheriton 1999], the group is specified in advance in the sense that the reference address of the group is explicitly included in the message. Groups or spaces in the ActorSpace model [Agha and Callsen 1993] are regarded as passive containers of processes (actors) and may be created and deleted with explicit constructs. Spaces may be nested or even overlap and can be created dynamically at run-time. Actually, the notion of space is a first class entity in the ActorSpace model. *AbC* extends the ActorSpace pattern-matching mechanism to select partners by predicates on both sides of the communication where not only the sender can select its partner but also the receiver can decide to either receive or discard the message. The notion of spaces/collectives in *AbC* is more abstract and is only specified at run-time.

In the *publish/subscribe* model, like the one used in NASA Goddard Mission Services Evolution Center (GMSEC)[5], each component can take the role of a publisher or a subscriber. Publishers produce messages and subscribers consume them. The subscribers are indirectly addressed by the contents of sent messages. That is, a subscriber expresses its interest independently from the publisher that produces the messages, and then it is asynchronously notified when a message that matches its interest arrives. The propagation of messages from publishers to subscribers is obtained by introducing an exchange server that mediates the interaction. The exchange server stores the subscriptions of subscribers, receives messages from publishers, and forwards the messages to the correct subscribers. The result is that publishers and subscribers are unaware of the existence of each other. Though the anonymity of interaction is an effective solution to achieve scalability by allowing participants to enter or leave the system at anytime, scalability issues are moved to the implementation of the exchange server. In fact, since the exchange server is responsible for subscriptions and message filtering, it should be able to face a large number of participants with evolving subscriptions while maintaining an acceptable level of performance.

As discussed before, the publish/subscribe model can be considered as a special case of *AbC* where publishers can attach attributes to messages and send them with empty predicates (i.e., satisfied by all). Only subscribers can check the compatibility of the attached publishers attributes with their subscriptions.

The Actor communication model, originally introduced in [Hewitt 1977; Hewitt et al. 1973], has been used to support the development of object-based concurrent computation. Actors embody the spirit of objects and extends their advantages to concurrent computations. As with objects where data and behavior are encapsulated to separate what an object can do from how it does it, the actors separate the control (where and when) from the logic of a computation. The early proposals of the Actor model in [Hewitt 1977; Hewitt et al. 1973] were rather informal. However, the definition of actors that is commonly used today follows Agha's definition [Agha 1986] that provides an abstract representation of actor systems in terms of what is called *Asynchronous Computation Trees*, building on notions borrowed from Milner's work [Milner 1989]. An actor system, also called a configuration, consists of autonomous objects, called actors, and of a collection of messages in transit. Computations in an actor system are carried out in response to received messages. Each message contains the destination identity and the actual content of the message. As opposed to *AbC* components, actors adopt binary communication which does not scale well in large systems. To interact, actors rely on identities; messages can only be sent to those actors whose identity is known. Thus actors can only interact if they know each other.

Programming collective and/or adaptive behavior has been studied in different research communities like those interested in Context-Oriented Programming (COP) [Hirschfeld et al. 2008] and in the Component-Based approach [Bruneton et al. 2004]. In Context-Oriented Programming (COP), a set of linguistic constructs is used to define context-dependent behavioural variations. These variations are expressed as partial definitions of modules that can be overridden at run-time to adapt to contextual information. They can be grouped via layers to be activated or deactivated together dynamically. These layers can be also composed according to some scoping constructs. Our approach is different in that components adapt their behaviour by considering the run-time changes of the values of their attributes which might be triggered by either contextual conditions or by local interaction. Another approach that considers behavioural variations is adopted in the Helena framework [Klarl 2015].

The component-based approach, represented by FRACTAL [Bruneton et al. 2004] and its Java implementation, JULIA [Bruneton et al. 2006], is an architecture-based approach that achieves adaptation by defining systems that are able to adapt their configurations to the contextual conditions. System components are allowed to manipulate their internal structure by adding, removing, or modifying connectors. However, in this approach interaction is still based on explicit connectors. In our approach predefined connections simply do not exist: we do not assume a specific architecture or containment relations between components. The connectivity is always subject to change at any time by means of attribute updates. In our view, *AbC* is definitely more adequate when highly dynamic environments have to be considered.

---

[5]http://opensource.gsfc.nasa.gov/projects/GMSEC_API_30/index.php

Finally, the attribute-based interaction has been exploited in the Carma specification language [Bortolussi et al. 2015; Loreti and Hillston 2016] which was designed based on the linguistic primitives of the *AbC* calculus to support quantitive analysis of collective-adaptive systems. Send and receive operations in Carma are associated with execution rates that can be dynamically adjusted based on environmental conditions. A Carma model can be automatically translated into a Continuous-Time Markov process that can be simulated and thus its dynamics can be studied quantitatively. *AbC* models are non-deterministic and they abstract from timing constraints. They are concerned with qualitative properties of the system being modelled. Clearly, *AbC* and Carma have been designed to serve different objectives in the sense that *AbC* is designed as a programming framework for CAS interactions while Carma is designed to quantitatively assess CAS models in terms of their performance.

*AbC* combines the lessons learnt from the above mentioned languages and calculi, in that it strives for expressiveness while aiming to preserve minimality and simplicity. The dynamic settings of attributes and the possibility of inspecting/modifying the environment gives *AbC* greater flexibility and expressiveness while keeping models as natural as possible.

## 6 CONCLUDING REMARKS

We have proposed a language-based approach for programming the interaction of collective-adaptive systems by relying on attribute-based communication. We have introduced the programming constructs of the *AbC* calculus and we exploited them to show how complex and interesting case studies, from the realm of collective-adaptive systems, can be programmed in an intuitive way. We illustrated the expressive power of attribute-based communication by showing the natural encoding of other existing communication paradigms. We argued that the general concept of attribute-based communication can be exploited to provide a unifying framework to encompass different communication models and interaction patterns. Since the focus of this article was to show the expressive power of attribute-based communication and their applicability in the context of CAS systems, we refrained from presenting theoretical results. However, full details about behavioural theory, equational laws, and a formal proof of encoding can be found in [Alrahman et al. 2016a, 2017a]. We also provided *AbC* APIs to enable attribute-based interaction in well-known programming languages, e.g., in Java [Alrahman et al. 2016b], in Go [Alrahman et al. 2017b], and in Erlang [De Nicola et al. 2017]. In [Alrahman 2017], we have also laid the basis for an efficient distributed implementation of the attribute-based interaction; we proposed three different communication infrastructures, proved their correctness, modelled them as stochastic processes, and finally evaluated their performance.

For future work, we want to investigate anonymity of interaction at the level of attribute identifiers. Indeed in our model, components are anonymous, but the "name-dependency" challenge arises at another level, that is, the attribute environments. In other words, the sender's predicate should be aware of the identifiers of receiver's attributes in order to explicitly use them. For instance, the sending predicate $loc = (1, 4)$ targets the components at location $(1, 4)$, however, different components might use different identifiers to denote their locations; this means that there should be Programming the Interactions of Collective Adaptive Systems 27an agreement about the attribute identifiers among components. For this reason, appropriate mechanisms for handling *attribute directories* together with identifiers matching/correspondence will be considered. These mechanisms will be particularly useful when integrating heterogeneous applications.

We plan to develop formal tools based on *AbC* semantics to analyse the code generated by our *AbC* APIs for ensuring safety and liveness properties. We want to study the possibility of using static analysis to discipline the interaction in *AbC* and thus producing a correct by construction programs and we will also consider the more challenging problem of specifying and verifying collective properties of *AbC* programs.

*Acknowledgments.* The authors would like to thank Francesco Tiezzi, Mirco Tribastone and Catia Trubiani for their valuable comments and suggestions to improve the quality of this article. We would also like to thank Giulio Garbi for debugging and implementing the case studies in GoAT.## REFERENCES

Gul Agha. 1986. *Actors: A Model of Concurrent Computation in Distributed Systems*. MIT Press, Cambridge, MA, USA.

Gul Agha and Christian J Callsen. 1993. *ActorSpace: an open distributed programming paradigm*. Vol. 28. ACM.

Yehia Abd Alrahman. 2017. A Foundational Theory for Attribute-based Communication. (2017). http://e-theses.imtlucca.it/id/eprint/214 PhD Thesis, IMT school for Advanced Studies Lucca.

Yehia Abd Alrahman, Rocco De Nicola, Giulio Garbi, and Michele Loreti. 2017b. Attribute-based Interaction in Google Go. https://github.com/giulio-garbi/goat. (2017).

Yehia Abd Alrahman, Rocco De Nicola, and Michele Loreti. 2016a. On the Power of Attribute-Based Communication. In *Formal Techniques for Distributed Objects, Components, and Systems - 36th IFIP WG 6.1 International Conference, FORTE 2016, Held as Part of the 11th International Federated Conference on Distributed Computing Techniques, DisCoTec 2016, Heraklion, Crete, Greece, June 6-9, 2016, Proceedings*. 1–18. https://doi.org/10.1007/978-3-319-39570-8_1

Yehia Abd Alrahman, Rocco De Nicola, and Michele Loreti. 2016b. Programming of CAS Systems by Relying on Attribute-Based Communication. In *Leveraging Applications of Formal Methods, Verification and Validation: Foundational Techniques - 7th International Symposium, ISoLA 2016, Imperial, Corfu, Greece, October 10-14, 2016, Proceedings, Part I*. Springer, 539–553. https://doi.org/10.1007/978-3-319-47166-2_38

Yehia Abd Alrahman, Rocco De Nicola, and Michele Loreti. 2017a. A Behavioural Theory for Interactions in Collective-Adaptive Systems. *ArXiv e-prints* (Nov. 2017). https://arxiv.org/abs/1711.09762

Yehia Abd Alrahman, Rocco De Nicola, Michele Loreti, Francesco Tiezzi, and Roberto Vigo. 2015. A Calculus for Attribute-based Communication. In *Proceedings of the 30th Annual ACM Symposium on Applied Computing (SAC '15)*. ACM, 1840–1845. https://doi.org/10.1145/2695664.2695668

S Anderson, N Bredeche, AE Eiben, G Kampis, and M van Steen. 2013. Adaptive collective systems: herding black sheep. *BookSprints for ICT Research* (2013).

Michael A Bass and Frank T Nguyen. 2002. Unified publish and subscribe paradigm for local and remote publishing destinations. (June 11 2002). US Patent 6,405,266.

Federico Bergenti, Giovanni Rimassa, and Mirko Viroli. 2003. Operational semantics for agents by iterated refinement. In *International Workshop on Declarative Agent Languages and Technologies*. Springer, 37–53.

Luca Bortolussi, Rocco De Nicola, Vashti Galpin, Stephen Gilmore, Jane Hillston, Diego Latella, Michele Loreti, and Mieke Massink. 2015. CARMA: Collective Adaptive Resource-sharing Markovian Agents. *Workshop on Quantitative Aspects of Programming Languages and Systems, QAPL 2015* (2015), 16–31.

Manfred Broy. 2006. The'grand challenge'in informatics: engineering software-intensive systems. *Computer* 39, 10 (2006), 72–80.

Eric Bruneton, Thierry Coupaye, Matthieu Leclercq, Vivien Quéma, and Jean-Bernard Stefani. 2006. The fractal component model and its support in Java. *Software: Practice and Experience* 36, 11-12 (2006), 1257–1284.

Eric Bruneton, Thierry Coupaye, and Jean-Bernard Stefani. 2004. The FRACTAL component model. *Draft of specification, version* 2, 3 (2004).

R. Calinescu. 2008. Implementation of a Generic Autonomic Framework. In *Fourth International Conference on Autonomic and Autonomous Systems (ICAS'08)*. 124–129. https://doi.org/10.1109/ICAS.2008.21

Betty HC Cheng, Rogerio De Lemos, Holger Giese, Paola Inverardi, Jeff Magee, Jesper Andersson, Basil Becker, Nelly Bencomo, Yuriy Brun, Bojan Cukic, et al. 2009. Software engineering for self-adaptive systems: A research roadmap. In *Software engineering for self-adaptive systems*. Springer, 1–26.

Gregory V Chockler, Idit Keidar, and Roman Vitenberg. 2001. Group communication specifications: a comprehensive study. *ACM Computing Survey (CSUR)* 33, 4 (2001), 427–469.

Frank S de Boer, Wieke de Vries, John-Jules Ch Meyer, Rogier M van Eijk, and Wiebe van der Hoek. 2005. Process algebra and constraint programming for modeling interactions in MAS. *Applicable Algebra in Engineering, Communication and Computing* 16, 2-3 (2005), 113–150.Manuscript submitted to ACM

# A APPENDIX

To provide further evidences of the expressive power of *AbC* and of its ability to naturally model the interaction in CAS, below we present two additional case studies from different application domains, namely crowd steering and swarm robotics.

## A.1 A Smart Conference System

This case study is inspired by work on crowd steering in public spaces during big events [Sassi et al. 2015]. The idea is to exploit the mobile devices of conference participants to guide them to the presentations they are interested in. Each participant expresses her topic of interest and the conference venue is responsible for guiding her to the location where the presentations that match her interest take place. The conference venue is composed of a set of rooms where the conference sessions are planned. We assume that the name of each room identifies its location, e.g., "$1^{st}$ Floor, Room.101" and that each participant has a unique *id*. The conference program and session relocation can be dynamically adjusted at anytime to handle specific situations, i.e., a crowded session can be moved to a larger room and this should be done seamlessly without any disruption to the whole conference program. When relocation happens, the updates should be communicated to the interested participants.

The smart conference system can be modelled in *AbC* as the parallel composition of the conference venue and the set of available participants. The conference venue is represented as a set of parallel *AbC* components, each of them representing a room ($Room_1 \| \ldots \| Room_n$) and each $Room_i$ has the following structure $\Gamma_i :_I R$ where $\Gamma_i$ represents the attribute environment of a room, $I = \{role, session\}$ represents its interface and $R$ represents its behaviour. A $Participant_i$ instead has the following structure $\Gamma_p :_{I'} P$ where $\Gamma_p$ represents its attribute environment, $I' = \{id, interest\}$ represents its interface, and $P$ represents its behaviour. The overall system is defined as follows:

$$Room_1 \| \ldots \| Room_n \| Participant_1 \| \ldots \| Participant_m$$

The attribute environments for participants and rooms contain the following attributes:

  **id**: the identity of a participant;
  **interest**: the current topic of interest of a participant;
  **name**: the name of a room, e.g., "$1^{st}$ Floor, Room.101";
  **dest**: to store the name of the room where the topic of interest for a participant is presented;
  **role**: the role of a component, each room has a *Provider* role;
  **session**: the current scheduled session for a room;
  **prevSession**: the session that was supposed to be held previously in a room if any;
  **newSession**: the new session assigned to a room;
  **relocate**: a boolean attribute, that indicates if a room has to start the relocation process.

The behaviour of process *P*, specifying the behaviour of a participant is defined as follows:

$$\textbf{let } v = (initialTopic) \textbf{ in } P \triangleq \textbf{set}(interest, v)$$
$$(\text{``request''}, \texttt{this}.id)@(role = \text{``Provider''}).$$
$$(session = \texttt{this}.interest \land x = \text{``interestRply''})(x, y).[dest := y]Upd$$

$$Upd \triangleq (x = \text{``update''} \land z = \texttt{this}.interest)(x, y, z, l).[dest := l]Upd$$

A participant starts by updating the value of her attribute *interest* with an initial topic. This will allow the participant to communicate her topic of interest to the conference venue by sending a session-request to nearby providers; in our case, this is a room. Notice that a participant does not know room names in advance, and only sends messages to components with a provider role. The communicated values include a label "*request*" and the identity of the participant. Once the message is emitted, the process blocks until a reply notification that matches her interest arrives. The notification contains an *interestRply* label, and the name of the room where the session is to be held. The participant is eligible to receive the message only if its attributes satisfy the sending predicate. Moreover, the participant receives only from a room with a session that matches her topic of interest (i.e., $session = \texttt{this}.interest$). By receiving this notification, the process updates its destination and waits for new possible updates through process $Upd$. This process is blocked until it receives an update notification about a session that matches the participant interest. The notification message contains an *update* label, the session that was supposed to be held previously in this room, the current session of the room, and the name of the room where the session of interest has been moved. Once a notification message is received, the process updates the destination to the new location and waits for future updates.

The behaviour of a room component is obtained as the parallel composition of three different processes:

$$R \triangleq Service \quad | \quad Relocation \quad | \quad Updating$$

The *Service* process is responsible for providing a normal service for a room. When a room receives a session-request, the process *Service* replies back to the requester with the room location whenever the current session of the room matches the topic of interest of the requester. The *Relocation* and the *Updating* processes are responsible for handling sessions and participants relocation from one room to another.

The process *Service*, reported below, shows how the room normally handles participant requests. The *Service* process is busy-waiting for session-requests from the available participants. Once a message with a possible session-request is received, and its attached interest value matches the current room session, the process sends an interest-reply message to the requester and gets ready to handle new session-requests by replicating itself. The interest-reply message contains an *interestRply* label and the name of the room. Notice that, since the predicate of the *interestRply* message addresses the requester with its identity, there is only one possible receiver for this message which is the requester herself.

$$Service \triangleq (x = \text{``request''} \land interest = \texttt{this}.session)(x, y).(\text{``interestRply''}, name)@(id = y).0 \;|$$
$$Service$$

The process *Relocation* handles unexpected changes of the schedule for a room. The idea is to handle these run-time changes in a way such that interested participants in the new session and also other rooms where a swap of schedule should happen are notified. The behaviour of *Relocation* is triggered by changes in the environmental attributes *NewSession* and *relocate*. In general, environmental attributes are controlled either by the environment itself or by other components; these may be humans or sensors that intervene to adapt the system behaviour to guarantee the expected functioning.

A possible scenario is that the session is becoming too crowded and needs to be relocated to another larger room with possibly few attendees. Another component that plays a portal role and keeps information about the capacity and the run-time utilisation of all rooms can propose a new session that best fits with the capacity of this room. Note that the portal component only proposes suitable sessions for rooms based on their capacities and their run-time utilisation, but has no control on their behaviours. The decision of relocation is made by the room itself depending on the readings from sensors which set the value of the attribute relocate to true if the level of overcrowding exceeded a certain threshold. To achieve relocation we have to steer the crowd from one room to another and vice versa. Actually, three different parties have to be notified about the changes of the schedule. The participants who are interested in the new proposed session, the room that is currently assigned the new proposed session and should swap its session with the crowded room, and the participants who are interested in the crowded session.

The *Relocation* process is responsible for notifying the room where a swap of schedule should happen and the participants who are interested in the new proposed session. The *Updating* process communicates the new location of the crowded session to the interested participants. An awareness operator is needed to allow the *Relocation* process to keep track of environmental changes; it is used to collect run-time data from the attribute environment of the component where the process *Relocation* is executed. These data are used to decide future behaviours.

Process *Relocation*, reported below, is blocked until the value of attribute *relocate* becomes true; it sends a session/interest update to the participants interested in the new assigned session and also to the room where a swap of schedule should happen as shown in the "∨" predicate of the send action. The interest update message contains an interest update label, the current session, the new session, and the name of the room. By sending this message, the process updates the room's previous session to the current one and the current session to the new session provided by the portal component through the value of the attribute *newSession*. The flag *relocate* is set to false.

$$Relocation \triangleq \langle relocate = \text{tt} \rangle (\text{``update''}, \text{this}.session, \text{this}.newSession, name)$$
$$@(interest = \text{this}.newSession \lor session = \text{this}.newSession).$$
$$[prevSession := session,$$
$$session := newSession, relocate := \text{ff}]Relocation$$

To relocate the sessions and steer participants correctly both the involved rooms do collaborate and propagate the changes in their schedules to the interested participants. The global goal is to make both groups of participants, interested in one of the two room sessions, aware of the new location of their interest. As seen before, process *Relocation* is partly responsible for propagating the changes to the group of participants interested in the newly assigned session. The same message is used to ask for collaboration to the other room, where a swap of schedule should happen. The other room has to update its session and propagate the changes to all participants interested in the crowded session. This task is performed by process *Updating*:

$$Updating \triangleq (x = \text{``update''} \land z = \text{this}.session)(x, y, z, l).$$
$$[prevSession := session, session := y]$$
$$(\text{``update''}, \text{this}.prevSession, \text{this}.session, \text{this}.name)$$
$$@(interest = \text{this}.session \lor session = \text{this}.session).0 \mid Updating$$

When a room receives a session-update message regarding its own session, the attribute *prevSession* gets the value of *session* and this is assigned the value of the session name communicated by the other room. Once the changes have been applied, the process sends a session/interest-update message and again the process *Updating* is made available to handle session-update messages. The session-update message is sent to the other group of participants so that they relocate to the new destination. It should be noted that the structure of this message is exactly the same as the one sent by *Relocation*. The only difference is that the sent values depend on the current attribute values of the room where the process *Updating* is executed.

*Example A.1 (Interaction fragment).* Let us assume that a participant, say $p_1$ with $id = 1$ and with "*Theory*" as her initial topic of interest, sent a session-request for nearby providers. The participant $p_1$ applies rule Comp twice. The first application of Comp, reported below, updates the *interest* attribute of the participant[6].

$$\Gamma_p :_{I'} P \xrightarrow{\{id=1,\ interest=\bot\} \triangleright \overline{\text{ff}}()} \Gamma_p[interest \mapsto \text{``Theory''}] :_{I'} P'$$

The second application of Comp is used to send the request to nearby providers:

$$\Gamma'_p :_{I'} P' \xrightarrow{\{id=1,\ interest=\text{``Theory''}\} \triangleright \overline{(role=\text{``Provider''})}(\text{``request''}, 1)} \Gamma'_p :_{I'} P''$$

If a room, say $r_1$ scheduled for the session "*Theory*", receives this message, it applies rule (**Comp**) and performs the following transition:

$$\Gamma_1 :_I Service \mid Relocation \mid Updating$$
$$\xrightarrow{\{id=1,\ interest=\text{``Theory''}\} \triangleright (role=\text{``Provider''})(\text{``request''}, 1)}$$
$$\Gamma_1 :_I Service'[\text{``request''}/x,\ 1/y] \mid Service \mid Relocation \mid Updating$$

All other rooms will just discard the request and apply rule FComp and the overall system, by applying rule ComR, will evolve as follows:

$$S \xrightarrow{\{id=1,\ interest=\text{``Theory''}\} \triangleright \overline{(role=\text{``Provider''})}(\text{``request''}, 1)}$$
$$(\Gamma_1 :_I Service'[\text{``request''}/x,\ 1/y] \mid Service \mid Relocation \mid Updating) \parallel$$
$$\Gamma'_p :_{I'} P'' \parallel \Gamma_2 :_I R_2 \parallel \ldots \parallel \Gamma_n :_I R_n$$

The components $\Gamma_2 :_I R_2, \ldots, \Gamma_n :_I R_n$ represent the other rooms in the conference venue. □

## A.2 A swarm robotics scenario in *AbC*

We consider a scenario where a swarm of robots spreads throughout a given disaster area with the goal of locating and rescuing possible victims. Similar case studies have been considered in [Ercan et al. 2006]. All robots execute the same code that defines their functional behaviour, and rely on a set of adaptation mechanisms that regulates the interactions with other robots and with the environment. Initially, all robots play the explorer role to search for victims in the environment. Once a robot finds a victim, its role changes to "rescuer" and it sends information about the victim to nearby explorers. The collective (the swarm) starts forming in preparation of the rescuing procedure. As soon as another robot receives information about the victim, it changes its role to "helper" and joins the rescuers-collective. The rescuing procedure starts only when the collective formation is complete. During exploration, in case of a critical battery level, a robot enters a power saving mode until it is recharged.

This model exploits the fact that a process running on a robot can either read the values of attributes provided by its sensors or read and update other attributes in its environment. Reading the values of the attributes controlled by sensors provides information about the environment or about the current status of the robot. Thus, the model captures both *self-awareness* and *context-awareness*. For instance, when reading the value of the *collision* attribute, if $\Gamma(collision) = \text{tt}$, the robot becomes aware that a collision with a wall in the arena is imminent and this triggers an adaptation mechanism

---
[6]Please remember that we have **set**($interest, InitialTopic$) ≡ ()@ff.[$interest := initialTopic$]

to change its direction. On the other hand, if $\Gamma(batteryLevel) = 15\%$, the robot becomes aware that its power is limited (e.g., < 20%) and starts an adaptation program to take the robot into the power saving mode.

We assume that each robot has a unique identity (*id*) and since the robot acquires information about its environment or its own status by reading the values provided by sensors, no additional assumption about its initial state is needed. It is worth mentioning that sensors and actuators are not modelled by *AbC* because they represent the robot internal infrastructure while the *AbC* model represents the programmable behaviour of the robot (i.e., its running code).

The robotics scenario is modelled as a set of parallel *AbC* components, each of which represents a robot ($Robot_1 \| \ldots \| Robot_n$) and each robot has the following form ($\Gamma_i :_I P_R$) where $I = \{id, role\}$.

The attribute environment for a robot contains the following attributes:

**id**: the unique identity of a robot;
**role**: the current role of a robot;
**victimPerceived**: to indicate if a victim is found;
**state**: its value triggers an actuation signal to either move or stop the movement;
**vPostion**: to indicate a victim position;
**count**: to indicate the required number of robots needed to rescue a victim;
**target**: its value triggers the actuators to move to a specific location;
**direction**: its value triggers the actuators to follow a specific direction;
**collision**: to indicate if a collision is detected;
**batteryLevel**: to indicate the battery level of the robot [7].

The behaviour of a single robot is modelled by:

$$P_R \triangleq (Rescuer + Explorer) | RandWalk | IsMoving$$

The robot follows a random walk in exploring the disaster arena. The robot can either become a "rescuer" when he becomes aware of the presence of a victim by locally reading the value of an attribute controlled by its sensors or remain as "explorer" and keep sending queries for information about the victim from nearby robots whose role is either "*rescuer*" or "*helper*".

The definition of process *Rescuer* is reported below. Please notice that the actual values of $x_1, x_2$ and $x_3$ are provided by external sensors.

$Rescuer \triangleq$
$\quad \langle \texttt{this}.victimPerceived = \texttt{tt} \rangle()@\texttt{ff}.[\texttt{this}.state := stop, \texttt{this}.vPosition :=  < x_1, x_2 >,$
$\quad\quad \texttt{this}.count := x_3, \texttt{this}.role := rescuer]$
$\quad (x = \text{"}qry\text{"} \wedge role = \text{"}explorer\text{"})(x, y).$
$\quad (\text{"}ack\text{"}, \texttt{this}.vPosition, \texttt{this}.count)@(id = y).0$

If sensors recognise the presence of a victim ("*victim-Perceived*" becomes "*tt*"), the robot updates its "*state*" to "*stop*" (which triggers an actuation signal to halt the actuators and stop movement), computes the position of the victim and the number of robots necessary to rescue the victim and stores them in "*vPosition*" and "*count*", then it changes its role to "*rescuer*" and waits for queries from nearby explorers. Once a message from an explorer is received, the robot sends back the information about the victim to the requesting robot addressing it with its identity "*id*".

---
[7]Notice that the values of attributes **victimPerceived, vPosition, count, collision,** and **batteryLevel** are provided by sensors while the values of attributes **state, target,** and **direction** send actuation signals to the actuators of the robot.

Alternatively, the robot plays the explorer role and continuously sends queries for information about the victim to nearby robots whose role is either "*rescuer*" or "*helper*". The query message contains a label "*qry*" to indicate the request type and the identity of the robot "this.*id*". When it receives an acknowledgement with victim's information, the robot changes its role to "*helper*" and starts the helping procedure. Notice that if an acknowledgment message is not received, the robot can again behave as rescuer or explorer as described above.

$$\begin{aligned}Explorer \triangleq{}& \\ & (\text{``qry''}, \text{this}.id)@(role = \text{``rescuer''} \vee role = \text{``helper''}). \\ & \quad (((role = \text{``rescuer''} \vee role = \text{``helper''}) \wedge x = \text{``ack''})(x, vpos, c).[\text{this}.role := helper]Helper \\ & \quad\quad + Rescuer \\ & \quad\quad + Explorer)\end{aligned}$$

The "*Helper*" process is reported below:

$$\begin{aligned}Helper \triangleq{}& ()@\text{ff}.[\text{this}.vPosition := vpos, \text{this}.target := vpos] \\ & \quad ( \ \langle \text{this}.position = \text{this}.target \rangle \mathbf{set}(\text{this}.role, rescuer)0 \\ & \quad\quad | \\ & \quad\quad \langle c > 1 \rangle (x = \text{``qry''} \wedge role = \text{``explorer''})(x, y). \\ & \quad\quad (\text{``ack''}, \text{this}.vPosition, c - 1)@(id = y).0 \ )\end{aligned}$$

The helping robot stores the victim position in the attribute "*vPosition*" and updates its target to be the victim position to tell the actuators to move to the specified location. While moving towards a victim, the robot is ready to respond to other robots queries, in case more than one robot ($c > 1$) is needed for the rescuing procedure. Once it reaches the victim (i.e., its position coincides with the victim position), the robot changes its role to "*rescuer*" and joins the rescuer-collective.

The "*RandWalk*" process is defined below, it determines the random direction to be followed by the robot. When a collision is detected by a proximity sensor, a new random direction is computed.

$$RandWalk \triangleq \mathbf{set}(\text{this}.direction, 2\pi rand())\langle \text{this}.collision = \text{tt} \rangle RandWalk$$

Finally, process "*IsMoving*" captures the status of the battery level in a robot at any time. Once the battery level drops into a critical level ($\leq 20\%$), the robot changes its status to "*stop*" which results in halting the actuators and the robot enters the power saving mode. The robot stays in this mode until it is recharged ($\geq 90\%$), then it starts moving again.

$$\begin{aligned}IsMoving \triangleq{}& \langle \text{this}.state = move \wedge (\text{this}.batteryLevel \leq 20\%) \rangle \\ & \quad \mathbf{set}(\text{this}.state, stop)\langle \text{this}.batteryLevel \geq 90\% \rangle \\ & \quad \mathbf{set}(\text{this}.state, move)IsMoving\end{aligned}$$

For simplifying the presentation, in this scenario we are not modelling the charging task and assume that this task is accomplished according to some predefined procedure.

*Example A.2 (Interaction fragment).* Let us assume that the role of $Robot_1$ is "*rescuer*" and $Robot_2$ is "*explorer*". $Robot_2$ can send a query to nearby rescuing or helping robots (i.e., $Robot_1$) by using rule Comp and generate this transition:

$$Robot_2 \xrightarrow{\{id=2,role="explorer"\} \triangleright \overline{(role="rescuer" \vee role="helping")}("qry", 2)} \Gamma_2 :_I (Explorer' \mid P_3)$$

where $P_3 \equiv RandWalk \mid IsMoving$

On the other hand, $Robot_1$ can receive this query by using rule Comp and generate this transition:

$$Robot_1 \xrightarrow{\{id=2,role="explorer"\} \triangleright (role="rescuer" \vee role="helping")("qry", 2)} \Gamma_1' :_I (Rescuer'["qry"/x, 2/y] \mid P_3)$$

where $P_3 \equiv RandWalk \mid IsMoving$

Other robots which are not addressed by communication discard the message by applying rule FComp. Now the overall system evolves by applying rule ComR as follows:

$$S \xrightarrow{\{id=2,role="explorer"\} \triangleright \overline{(role="rescuer" \vee role="helping")}("qry", 2)}$$

$$\Gamma_1' :_I (Rescuer'["qry"/x, 2/y] \mid P_3) \parallel \Gamma_2 :_I (Explorer' \mid P_3) \parallel \Gamma_3 :_I P_{R_3} \parallel \ldots \parallel \Gamma_n :_I P_{R_n}$$

□